\documentclass{aastex}
\shorttitle{FIR LINE AND DUST EMISSION FROM H \footnotesize II \large\ REGIONS AND PDRs}
\shortauthors{J. Zhu & M. Huang}
\begin{document}
\title{FAR-INFRARED LINE AND DUST EMISSION FROM H \footnotesize II \large\ REGIONS AND PDRs}

\author{Jiali Zhu\altaffilmark{1} and M. Huang}
\affil{National Astronomical Observatories, Chinese Academy of Sciences, Beijing, 100012}
\email{mhuang@nao.cas.cn}
\altaffiltext{1}{University of Chinese Academy of Sciences, 100049, Beijing.}

\begin{abstract}
We explore the effect of varying the spectral energy distribution of incident continuum, by simultaneously and self-consistently computing the structure of an H \footnotesize II\normalsize\ region and a photodissociation region (PDR) that are in pressure equilibrium. The results of calculation are applied to extragalactic observations. The intensity ratio diagrams of far-infrared (FIR) emission for Herschel bands (70, 110, 160, 250, 350 and 500 \micron) and the contribution from H \footnotesize II\normalsize\ regions for these specific FIR emission are presented for the first time. With these diagrams, we compare the predicted FIR continuum intensity ratios of M82 with observations by Herschel.
\end{abstract}


\section{INTRODUCTION}

Photodissociation regions (PDRs; \citealt{TH85}) are regions of the interstellar medium (ISM) where far-ultraviolet (FUV; 6 eV$< h\nu< $13.6 eV) photons dominate the structure, chemistry and thermal balance of gas \citep{ht97}. All neutral atomic hydrogen gas and a large fraction of the molecular gas in galaxies are in PDRs. PDRs are the origin of most of the non-stellar infrared emission from galaxies, including far-infrared (FIR) continuum from dust grains, near and mid infrared (IR) emission from polycyclic aromatic hydrocarbons (PAH), as well as fine structure IR emission such as [O \footnotesize I\normalsize ] 63 \micron\ and 146 \micron, [C \footnotesize I\normalsize] 370 \micron\ and 609 \micron, [Si \footnotesize II\normalsize] 35 \micron\ and [C \footnotesize II\normalsize] 158 \micron. 

H \footnotesize II\normalsize\ regions adjacent to PDRs are known to contribute to line emission and FIR continuum that are also found in the surrounding PDRs. \citet{hei94} found that the ionized medium contributes to [C \footnotesize II\normalsize] 158 \micron\ line luminosity. [O \footnotesize I\normalsize] 63 \micron\ and 146 \micron, [Si \footnotesize II\normalsize] 35 \micron\ and [Fe \footnotesize II\normalsize] 26 \micron\ line emission also exist in H \footnotesize II\normalsize\ regions \citep{abel2005,kau2006}. Dust grains in H \footnotesize II\normalsize\ regions absorb ionizing photons and reemit in FIR continuum \citep{bot98}. Thus, when we observe the H \footnotesize II\normalsize\ region and PDR in one telescope beam, the FIR and line emission generally comes from both regions.

There are two methods to derive properties and contributions of H \footnotesize II\normalsize\ regions and PDRs separately. Both methods treat radiation processes of these two regions with distinctive differences. The first method is to use the [N \footnotesize II\normalsize] 122 \micron/[C \footnotesize II\normalsize] 158 \micron\ ratio \citep{hei94,mal2001} and [N \footnotesize II\normalsize] 205 \micron/[C \footnotesize II\normalsize] 158 \micron\ ratio \citep{obe2006} in ionized medium. Since nitrogen has a first ionization potential (14.5 eV) higher than that of Hydrogen, [N \footnotesize II\normalsize] is found only in H \footnotesize II\normalsize\ regions. Using the [N \footnotesize II\normalsize]/[C \footnotesize II\normalsize] ratio, one can derive the emission of [C\footnotesize~II\normalsize] 158~\micron\ that arises from H \footnotesize II\normalsize\ regions. This method is only useful for deriving [C \footnotesize II\normalsize] 158 \micron\ line emission. The second method is to calculate a separate model for each region. \citet{car94} estimated the  [C \footnotesize II\normalsize] 158~\micron\ emission of H \footnotesize II\normalsize\ regions from models of \citet{rub85} and the [C \footnotesize II\normalsize] 158 \micron\ emission of PDRs from models of \citet{TH85} and \citet{HTT91}. A similar approach is taken to estimate the contributions for [Si \footnotesize II\normalsize] 35 \micron\ line emission in M82 \citep{lor96}. \citet{col99} combined starburst H \footnotesize II\normalsize\ region models and PDR models of \citet{k99} to derive the H \footnotesize II\normalsize\ region and PDR properties of M82. \citet{kau2006} computed a separate model for each region. They merged the H \footnotesize II\normalsize\ region model and PDR model by equaling the thermal pressure at the interface. The edge of the H \footnotesize II\normalsize\ region is defined at the point where H is 50\% neutral. This kind of separated calculation must take great care to assure that the transmitted continuum emerging from the H \footnotesize II\normalsize\ region is consistent with the initial conditions for the PDR \citep{abel2006}. 

\section{THE A05 MODEL OF THE H \footnotesize II\normalsize\ REGION AND PDR}
Using a procedure different from the above, \citet{abel2005} self-consistently calculated the thermal and chemical structure of an H~\footnotesize II\normalsize\ region and a surrounding PDR that are in pressure equilibrium (henceforth the A05 model). In this method, they viewed the H~\footnotesize II\normalsize\ region and PDR as a single region driven by UV radiation from stars.  This treatment has been tested in various environments (e.g., \citealt{pel2007,pel2009, hen2007, hal2008, gra2011}). The A05 model produces diagnostics without needing to assume how much of the emission is from H~\footnotesize II\normalsize\ regions or from PDRs. The advantage of the A05 model is to shield the complexity of the boundaries between H~\footnotesize II\normalsize\ regions and PDRs, and provide observables based on parameterized stellar radiation, gas density, composition, and geometry.

In the A05 model, parameterized UV flux of stars is the source of ionization and photodissociation of the ISM, creating H \footnotesize II\normalsize\ regions and surrounding PDRs. The spectral energy distribution (SED) of stellar atmospheres influences the ionization structures of  H \footnotesize II\normalsize\ regions and PDRs. \citet{mor2004} computed models of  H \footnotesize II\normalsize\ regions using a variety of recent state-of-the-art stellar atmospheres models. They compared model predictions to catalogs of ISO observations of Galactic H \footnotesize II\normalsize\ regions, and confirmed the finding of earlier investigation showing that CoStar \citep{sch97} atmospheres adopted by \citet{abel2005} over-predict somewhat the ionizing flux at high energies. They also concluded that WMBasic \citep{pau2001} atmospheres show a reasonable agreement with the observations. 

In this paper, we adopt WMBasic stellar atmospheres, repeat the calculations presented in \citet{abel2005}, using the same dynamical range for ionization parameter, density, equation of state, and abundances, and extend wavelength coverage to Herschel\footnote{{\it Herschel} is an ESA space observatory with science instruments provided by European-led Principal Investigator consortia and with important participation from NASA.} FIR bands up to 500 \micron. Then we apply our results to M82 and NGC~253. At the same time, we explore the effect of varying stellar atmospheres in the A05 model, and compare our work with \citet{abel2005}. We perform calculations for CoStar and WMBasic atmospheres at effective temperature $T=34000$ K and at $T=38000$~K. The model calculations are presented in Section 3, and the results are shown in Section 4. In Section 5, we compare model predictions to observational data in literature. We summarize in Section 6.

\section{MODEL CALCULATIONS}

The calculation details are the same as \citet{abel2005}. The Cloudy\footnote{version 13.02} code last described by \citet{fer2013} is used in calculation. We also define the end of the H \footnotesize II\normalsize\ region in the same way as Abel et al. (2005). The major differences between the \citet{abel2005} calculations and the ones presented here are that we use the WMBasic stellar radiation field, and the calculations are stopped at A$_{\rm{v}}=10$ instead of 100.

\begin{figure}
\includegraphics[scale=.70]{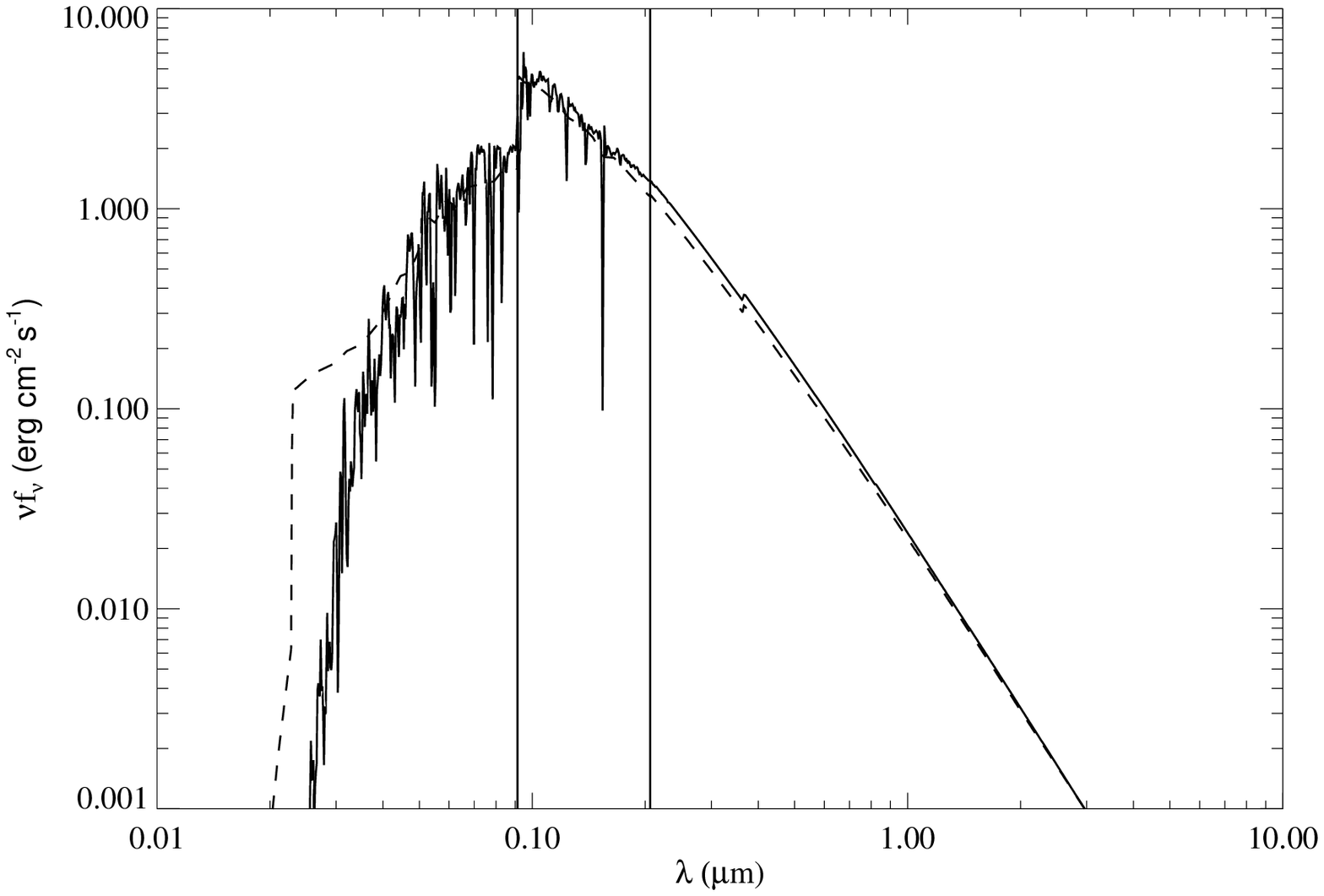}
\caption{Incident continuum profile, with $\log U=-2$, $\log n$(H)$=2$ cm$^{-3}$, and $T=38000$~K. The solid line is WMBasic continuum and the dash line is CoStar continuum. The area between two vertical lines is the continuum used to define $G_0$. The ionizing flux of CoStar atmospheres is higher than that of WMBasic atmospheres. \label{fig1}}
\end{figure}

In order to explore the effect of different stellar atmospheres, we compute models for WMBasic and CoStar atmospheres with incident continuum as shown in Figure \ref{fig1}. Figure \ref{fig1} shows that the CoStar radiation field produces more hydrogen ionizing flux than WMBasic.

\begin{figure}
\includegraphics[scale=.70]{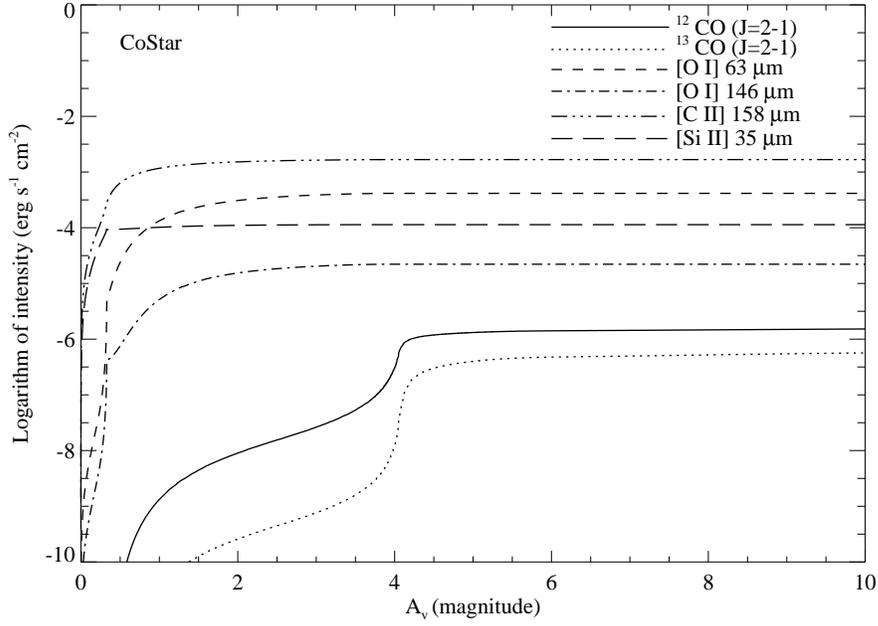}
\includegraphics[scale=.70]{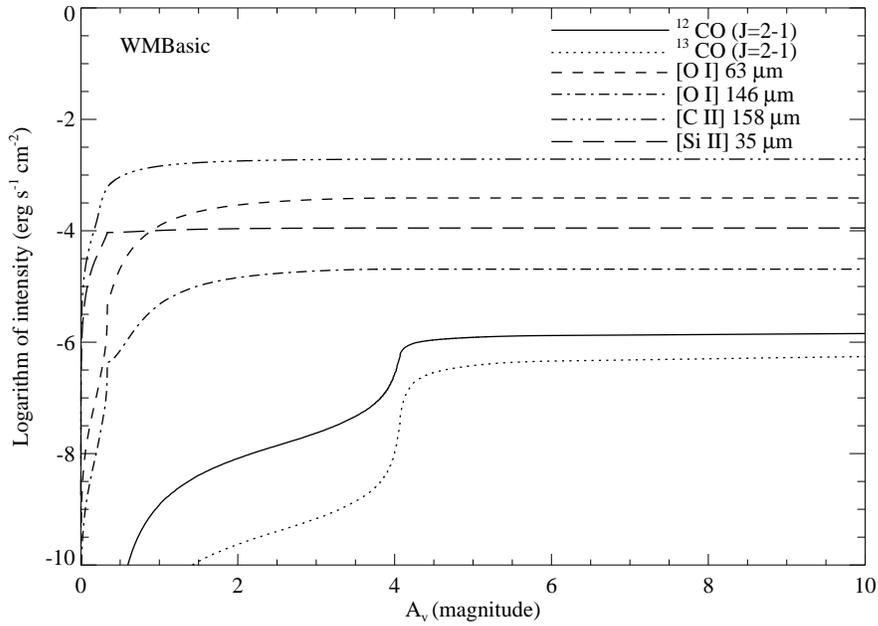}
\caption{Integrated intensity of PDR lines. Intensity of lines is stable at A$_{\rm{v}}=10$, while $\log U=-2$, $\log n$(H)$=1$ cm$^{-3}$, and $T=38000$ K.\label{fig2}}
\end{figure}

In the model, the H \footnotesize II\normalsize\ region and the PDR are placed between the ionizing source and the observers. Thus, we observe the transmitted continuum and outward emission from the emitting cloud. We define that the PDR extends to a visual extinction A$_{\rm{v}}=10$. At that depth, hydrogen is molecular and carbon is incorporated into CO. Figure \ref{fig2} shows that the integrated intensity of PDR lines is stable at A$_{\rm{v}}=10$. 

\section{RESULTS}

In this section we present the calculation results in a series of contour plots for all the $U$, $n$(H) and stellar atmospheres. Diagnostic diagrams similar to ones in \citet{abel2005} are not presented here, since they are insensitive to the choice of steller continuum. We show the differences between CoStar and WMBasic atmospheres for the A05 model in Section 4.1. Intensity ratios of FIR emission for the 70, 110, 160, 250, 350 and 500 \micron\ are first presented in Section 4.2.

\subsection{Differences between the CoStar and WMBasic Atmospheres for the A05 Model}
The strength of ionizing radiation field can be constrained by the intensity ratio of emission lines from sequential stages of ionization of a single element. [Ne  \footnotesize III\normalsize]  15.5 \micron/[Ne \footnotesize II\normalsize] 12.8 \micron\ ratio (Figure \ref{fig3}) and [S  \footnotesize IV\normalsize] 10.5 \micron/[S  \footnotesize III\normalsize] 18.7 \micron\ ratio (Figure \ref{fig4}) are good measures of the hardness of the radiation field \citep{bei2008}. [Ne \footnotesize III\normalsize ] 15.5 \micron/[Ne \footnotesize II\normalsize] 12.8 \micron\ ratio plots are more horizontal than  [S  \footnotesize IV\normalsize ] 10.5 \micron/[S  \footnotesize III\normalsize] 18.7 \micron\ ratio plots. Comparing the plots for WMBasic and CoStar atmospheres, we find that at $T=38000$ K [Ne \footnotesize III\normalsize ] 15.5~\micron/[Ne \footnotesize II\normalsize] 12.8 \micron\ ratio for Costar atmospheres is 50 times greater than that for WMBasic atmospheres, while [S  \footnotesize IV\normalsize ] 10.5 \micron/[S  \footnotesize III\normalsize] 18.7 \micron\ ratio for CoStar atmospheres is 10 times greater than that for WMBasic atmospheres. With the same $U$, WMBasic atmospheres need a higher stellar temperature than CoStar atmospheres to produce the same ratios. The results that these H \footnotesize II\normalsize\ region lines are sensitive to the ionizing flux distribution was also shown in \citet{mor2004}.

Figure \ref{fig1} shows the FUV continuum is nearly identical between CoStar and WMBasic atmospheres.  As a result, our calculations for the $G_0$ (in units of the local Galactic FUV flux$=1.6\times10^{-3}$ ergs cm$^{-2}$ s$^{-1}$; \citealt{hab68}), density as a function of depth, the PDR line ratios, and the contribution of traditional PDR lines except for the [C \footnotesize II\normalsize] 158 \micron\ line from the H \footnotesize II\normalsize\ region are essentially unchanged between \citet{abel2005} and this work. Density diagnostics, [O \footnotesize III\normalsize] 52 \micron/[O \footnotesize III\normalsize] 88 \micron\ ratio and [S \footnotesize III\normalsize] 18.7 \micron/[S \footnotesize III\normalsize] 33.5 \micron\ ratio, are not sensitive to stellar atmospheres. Therefore we do not present these diagrams in this work and refer to the \citet{abel2005} results in application.

The Figure \ref{fig5} shows the difference in the contribution for [C \footnotesize II\normalsize] 158 \micron\ between the two stellar atmospheres. This kind of difference has been found by \citet{abel2006}, who compared Kurucz \citep{kur79} stellar atmopsheres along with WMBasic and a blackbody in analyzing the  [C \footnotesize II\normalsize] contribution from the ionized gas.

\subsection{FIR Thermal Dust Emission} 
Interstellar dust in galaxies absorbs energy from starlight and re-radiates at IR and FIR wave range. PACS \citep{pog2010} and SPIRE \citep{gri2010} onboard Herschel \citep{pil2010} observe at 70, 110, 160, 250, 350 and 500~\micron. Ratios of emission in these wavebands indicate the dust temperature and brightness (e.g., \citealt{rou2010}).

Abel et al. (2009) showed the 60 \micron/100 \micron\ ratio, and the fraction of total FIR emitted by dust at the H \footnotesize II\normalsize\ ionization front. Here we present the first calculations of the FIR continuum ratios for the Herschel bands (Figure \ref{fig6} and \ref{fig7}). Figure 6 and Figure 7 show that the CoStar and WMBasic plots give essentially the same FIR band ratios. The contribution to FIR continuum emission from H \footnotesize II\normalsize\ regions is the same for either stellar atmospheres and therefore we only show the results for the WMBasic model in Figure \ref{fig8}.

At relatively low density, the contributions from H \footnotesize II\normalsize\ regions for 350 and 500 \micron\ emission depend strongly on $U$ rather than on density, and H \footnotesize II\normalsize\ region contributes more at higher $U$. At the upper-right corner of contour plots for 350 and 500 \micron\, where the density and $U$ are high, contributions depend both on $U$ and density. For 160 and 250~\micron\ emission,  the contributions from H \footnotesize II\normalsize\ regions depend strongly on $U$ rather than on density. For 70 and 110 \micron\ emission,  the contributions from H \footnotesize II\normalsize\ regions depend both on $U$ and density. PDR is the main origin of 110, 160, 250 and 350~\micron\ continuum emission, although H \footnotesize II\normalsize\ regions still contribute more than 20\%  emission at $\log U>-1.5$. H~\footnotesize II\normalsize\ regions can dominate the 500 \micron\ emission when both density and $U$ are high ($\log U>-2$ and $\log n$(H)$>3$ cm$^{-3}$), and dominate the 70 \micron\ emission when $\log U>-2$ and $\log n$(H)$<2.5$ cm$^{-3}$. The H \footnotesize II\normalsize\ region contributes more at higher $U$ because a larger fraction of the UV and ionizing photons are absorbed by dust in the H \footnotesize II\normalsize\ region.

Going to a higher Av will cause colder dust to affect the overall observed FIR emission. To give some insight into this effect, we calculate the A05 model at A$_{\rm{v}}=5$, 10, 50, 100, and 200 for $\log U=-2$ and $\log n$(H)$=2$ cm$^{-3}$. The predicted temperature of graphite with size 0.1 \micron\ is 21 K at A$_{\rm{v}}=5$, 18 K at A$_{\rm{v}}=10$, 16 K at A$_{\rm{v}}=50$ and A$_{\rm{v}}=100$, and 15 K at A$_{\rm{v}}=200$. 

\section{APPLICATION TO EXTRA-GALAXIES}

We apply results to extra-galaxies, and explore the influence of different stellar atmospheres. Comparing observations with our models, we derive $U$ (Figure \ref{fig3} and \ref{fig4}). Comparing observations with Figure 22 of \citet{abel2005},  we derive $n$(H). Then comparing Figure \ref{fig5}, \ref{fig7} and \ref{fig8} in this work and Figure 16, 17, 27, 29 and 33 of \citet{abel2005} with the derived $U$ and $n(H)$, we can derive other parameters.

To compare our results with \citet{abel2005} and the separated treatment of H \footnotesize II\normalsize\ regions and PDRs \citep{car94}, we apply our results to NGC~253, which was also analyzed by \citet{abel2005} and \citet{car94}.  

M82 is observed by Herschel recently to obtain FIR continuum flux at 250, 350 and 500~\micron\ \citep{rou2010}. We use our results for FIR continuum (Figure \ref{fig7} and \ref{fig8}) to predict properties of dust emission in H \footnotesize II\normalsize\ regions and PDRs, and compare it with observations.

\subsection{Application to NGC~253 and Comparison with \citet{abel2005}} 

\begin{figure}
\includegraphics[scale=.55]{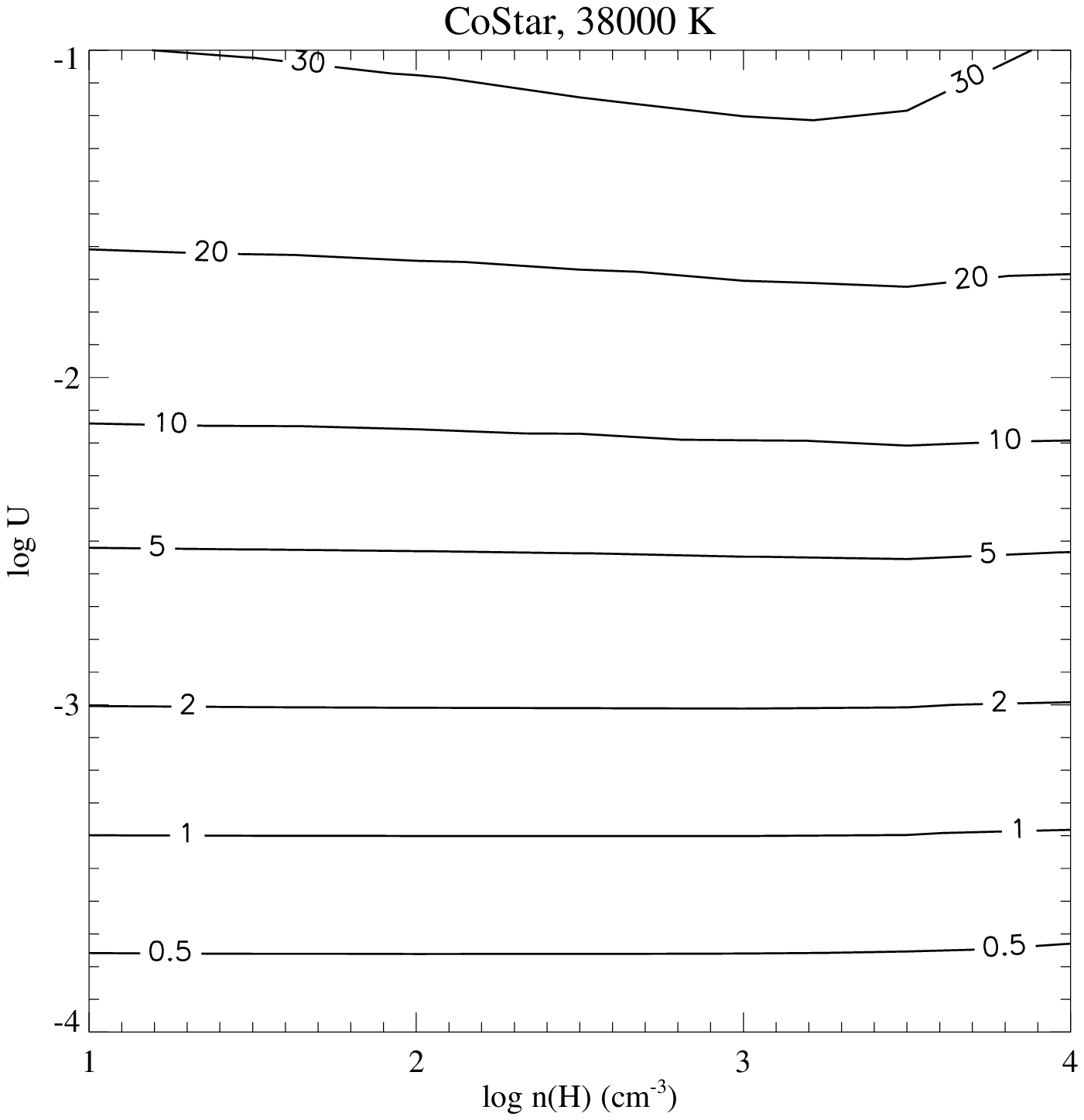}
\includegraphics[scale=.55]{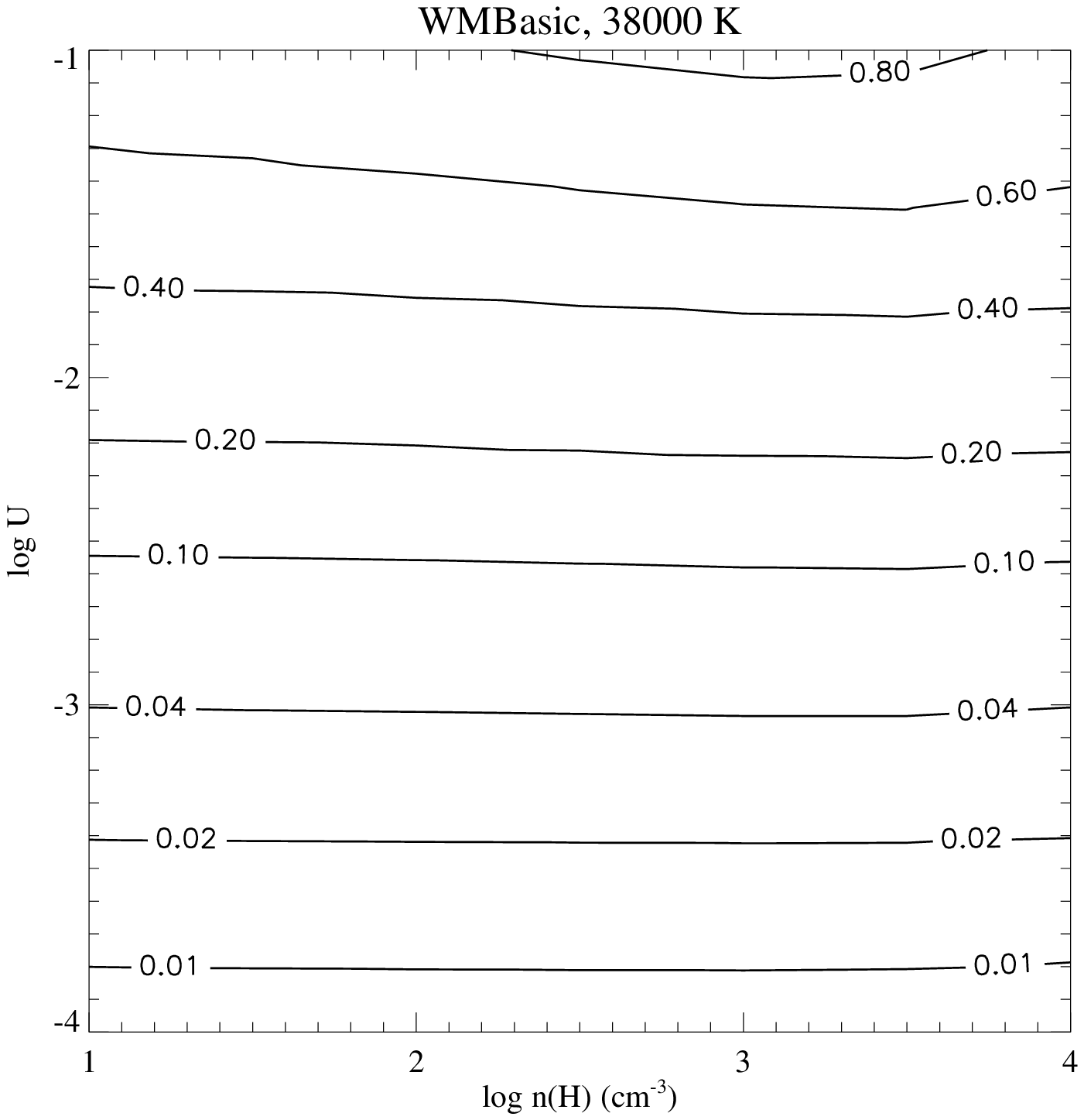}
\includegraphics[scale=.55]{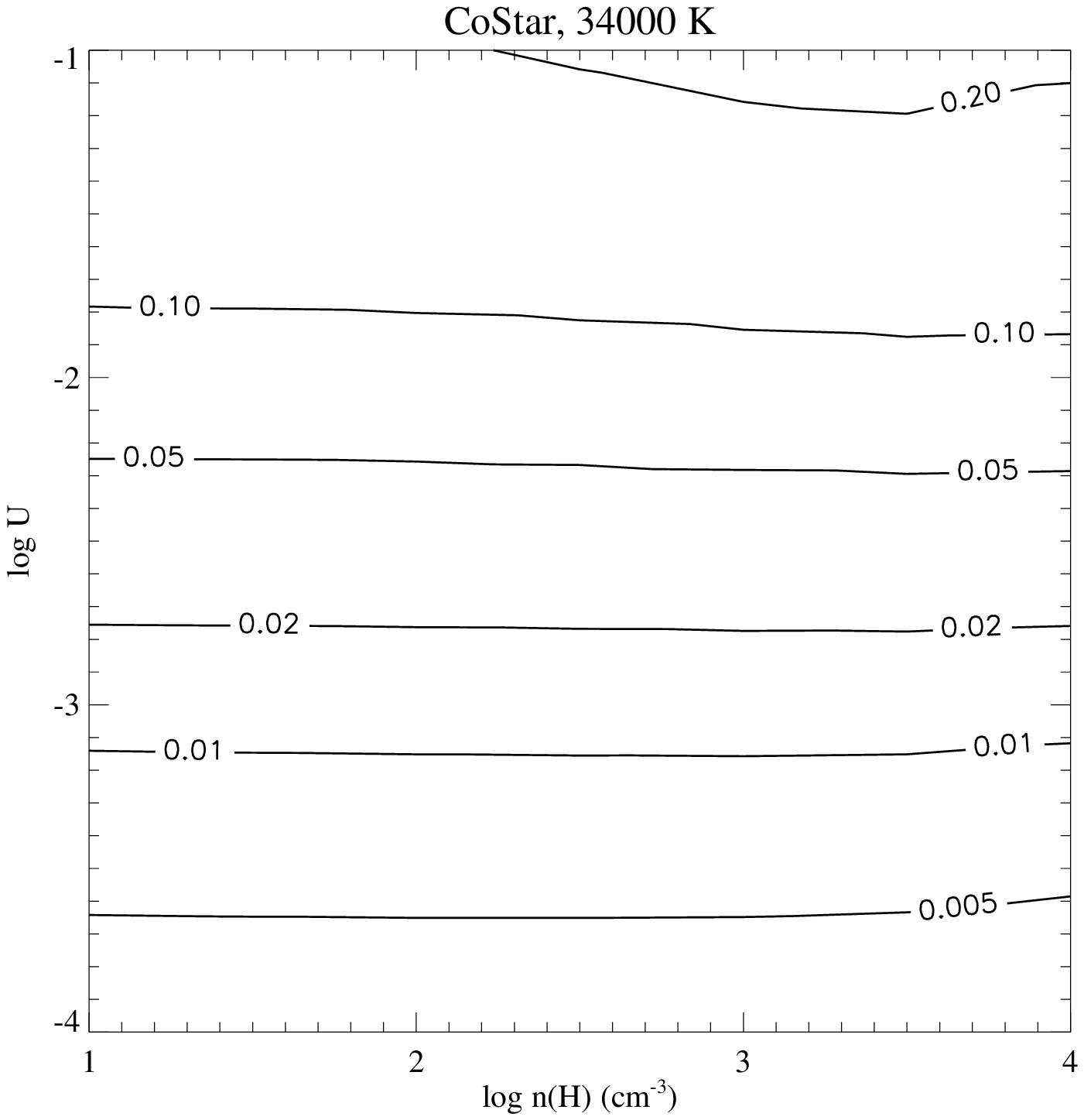}
\includegraphics[scale=.55]{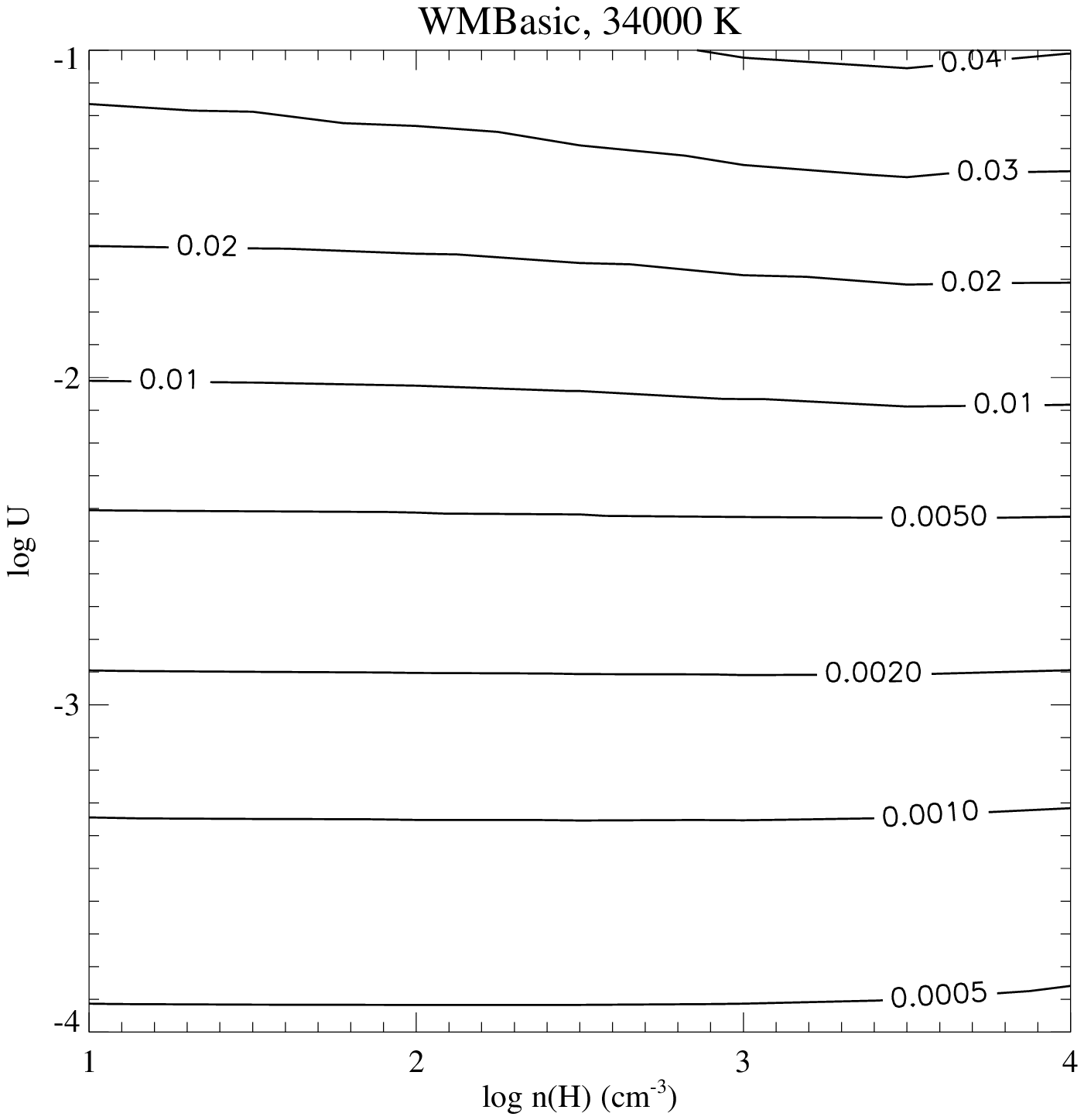}
\caption{[Ne \footnotesize III\normalsize ] 15.5 \micron/[Ne \footnotesize II\normalsize ] 12.8 \micron\ intensity ratio for WMBasic and CoStar atmospheres. \label{fig3}}
\end{figure}

\begin{figure}
\includegraphics[scale=.55]{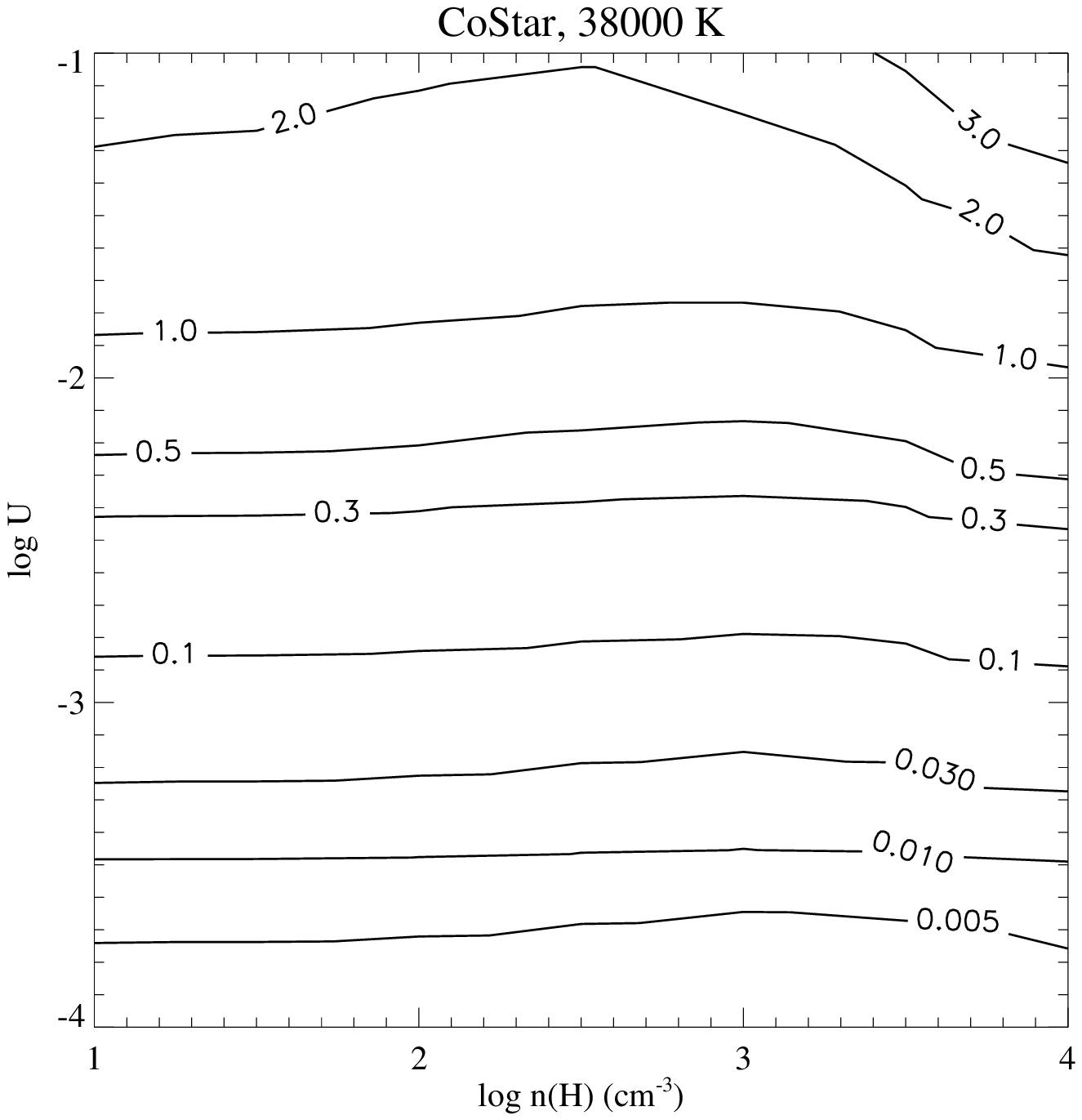}
\includegraphics[scale=.55]{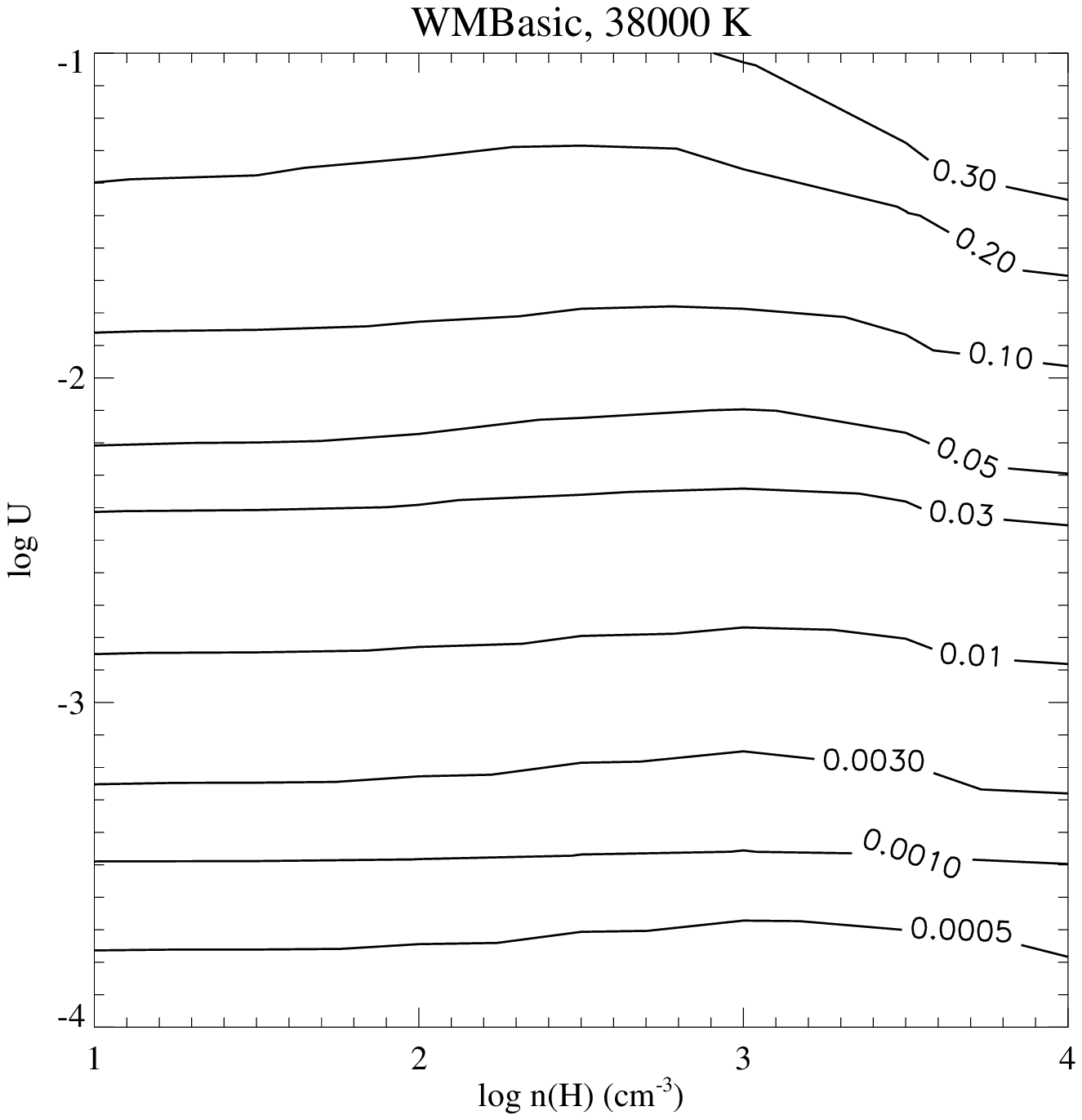}
\includegraphics[scale=.55]{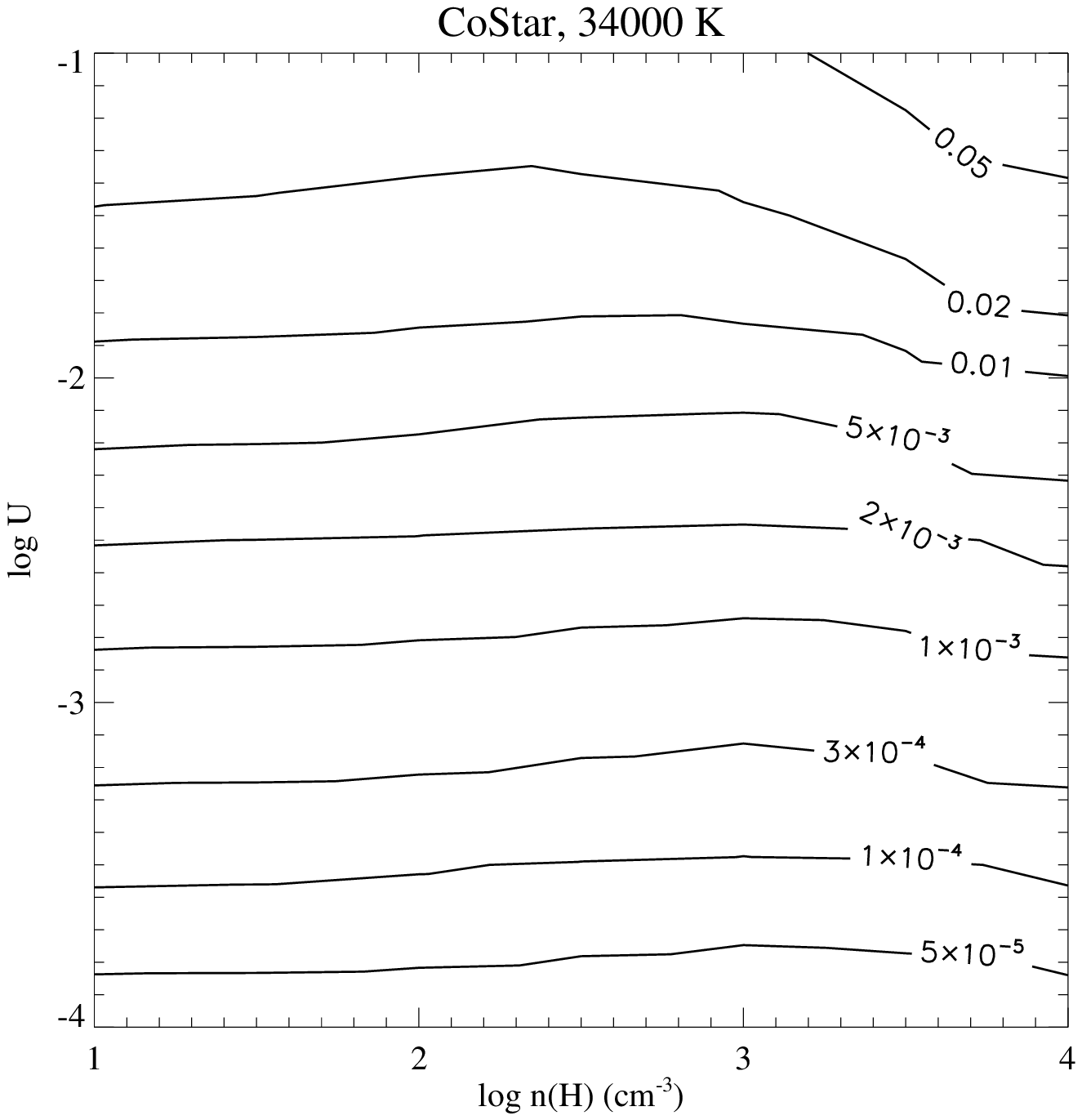}
\includegraphics[scale=.55]{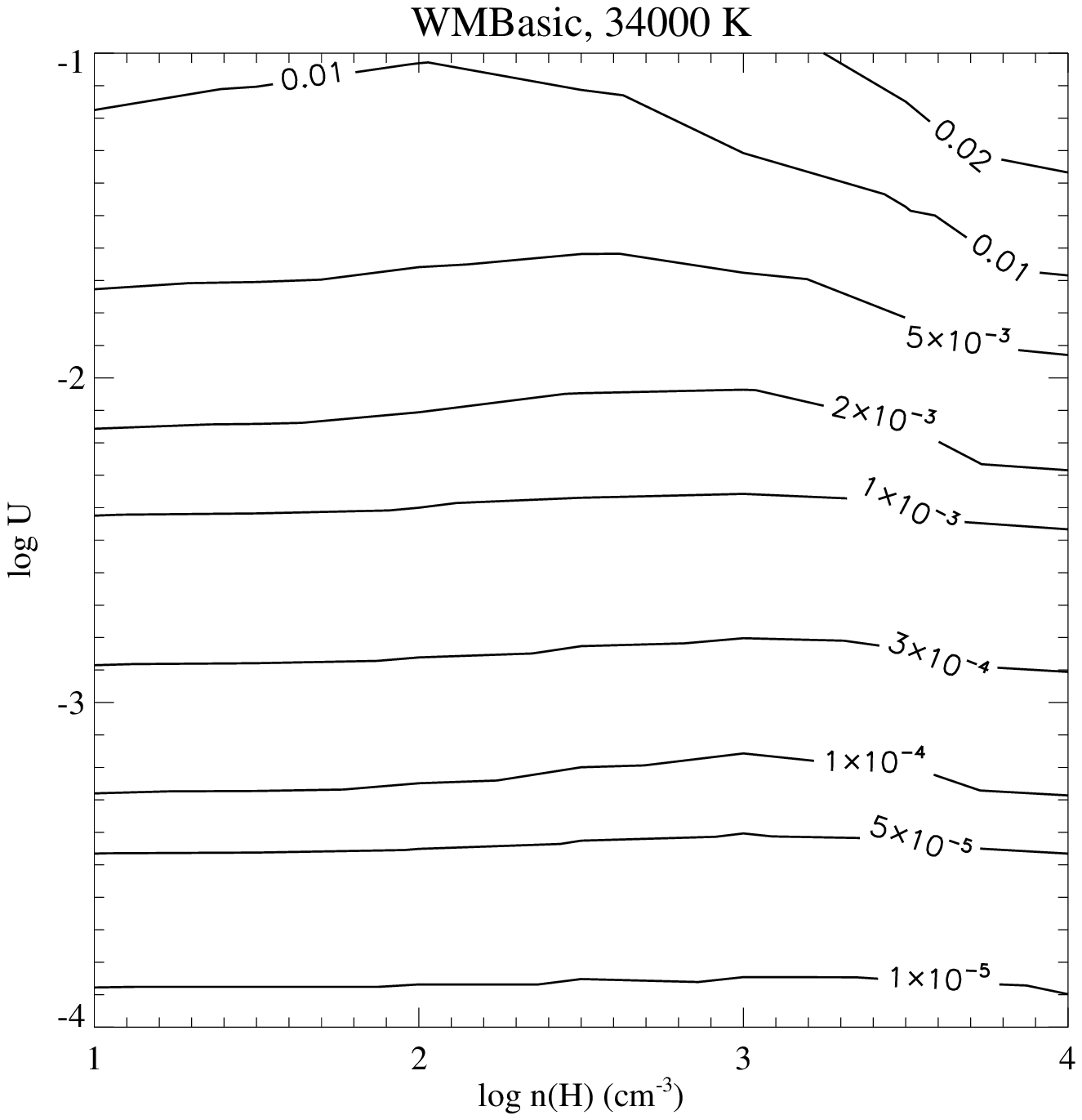}
\caption{[S  \footnotesize IV\normalsize ] 10.5 \micron/[S  \footnotesize III\normalsize] 18.7 \micron\ intensity ratio for WMBasic and CoStar atmospheres.\label{fig4}}
\end{figure}

\begin{figure}
\includegraphics[scale=.55]{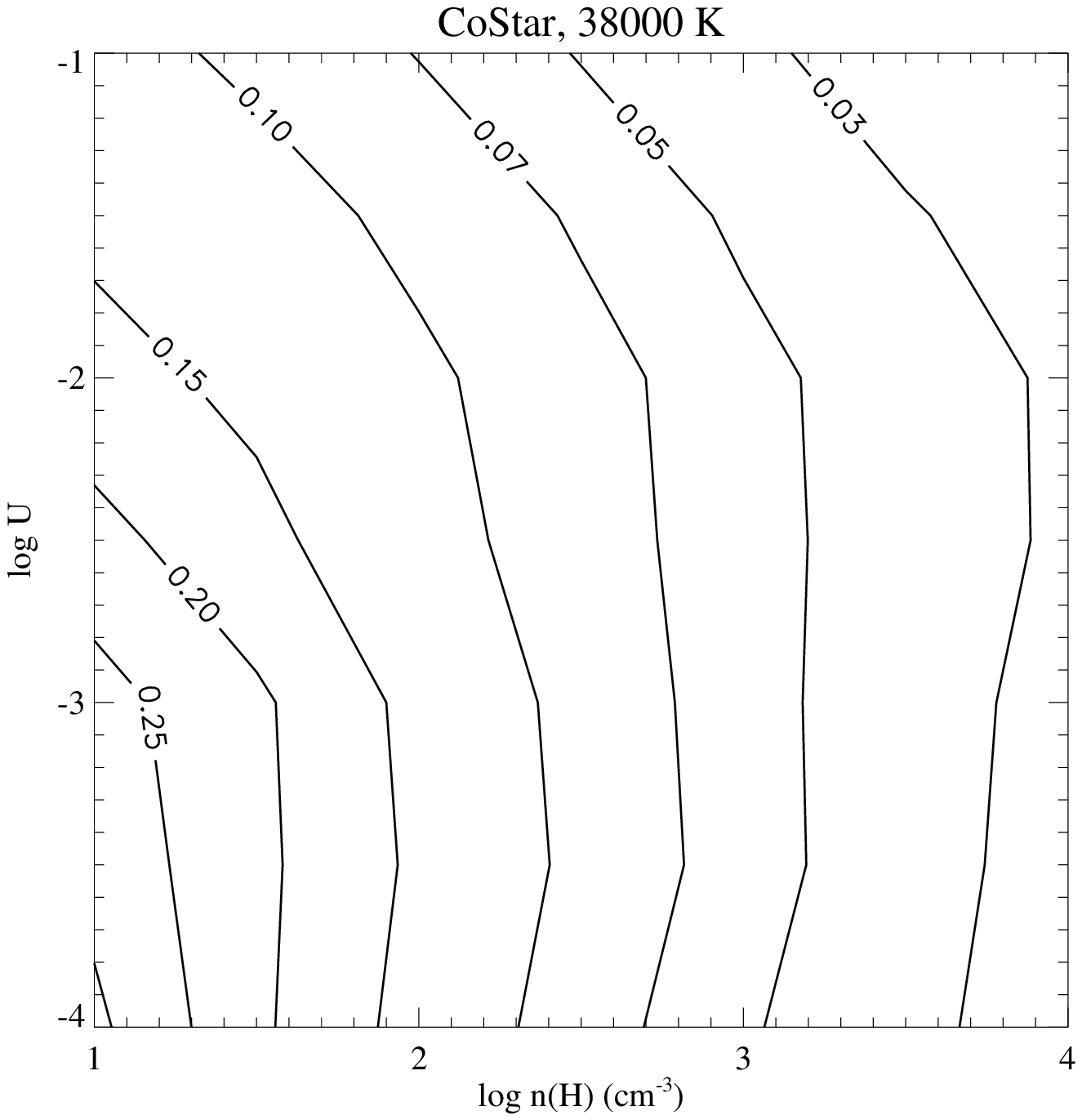}
\includegraphics[scale=.55]{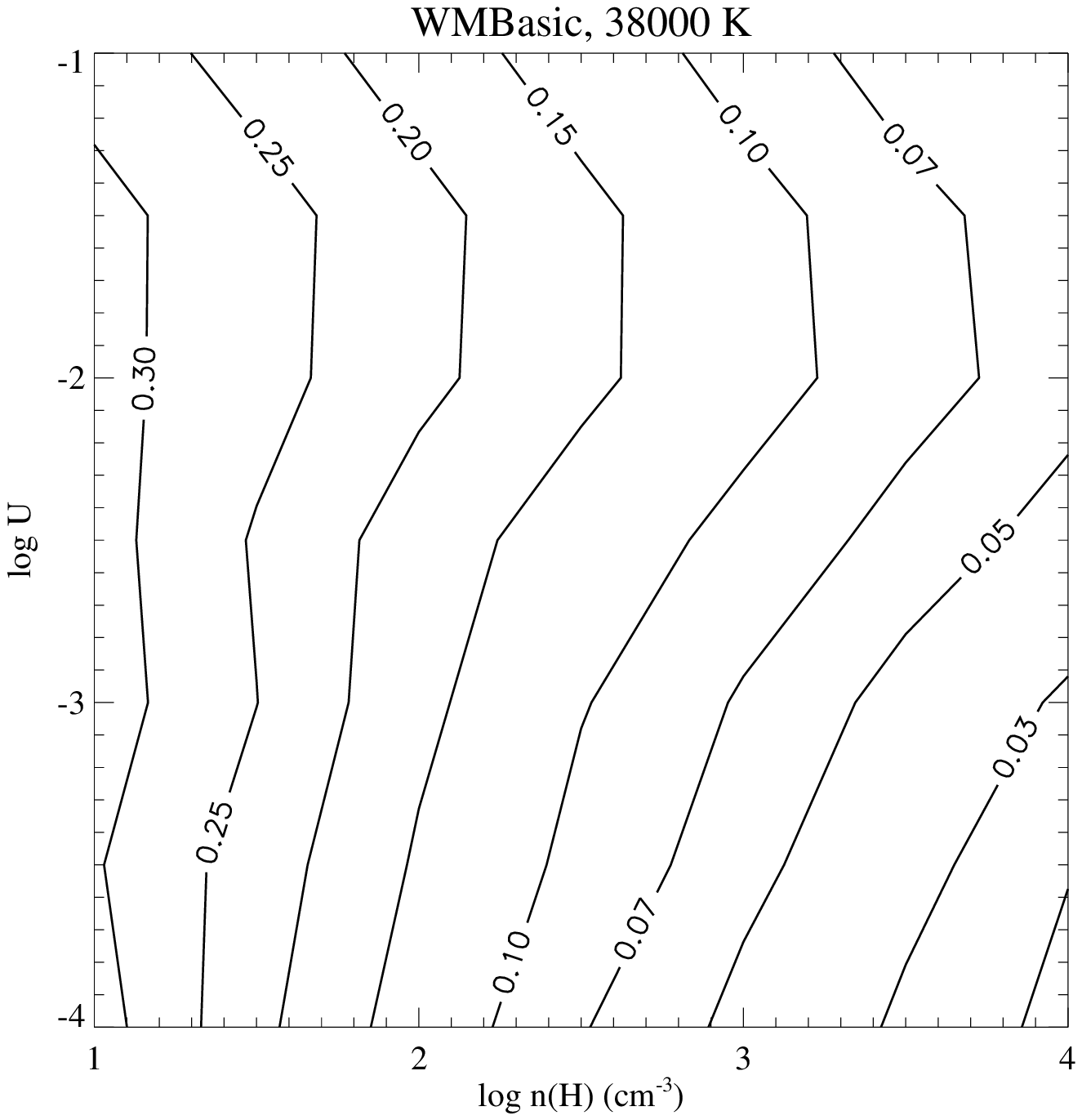}
\includegraphics[scale=.55]{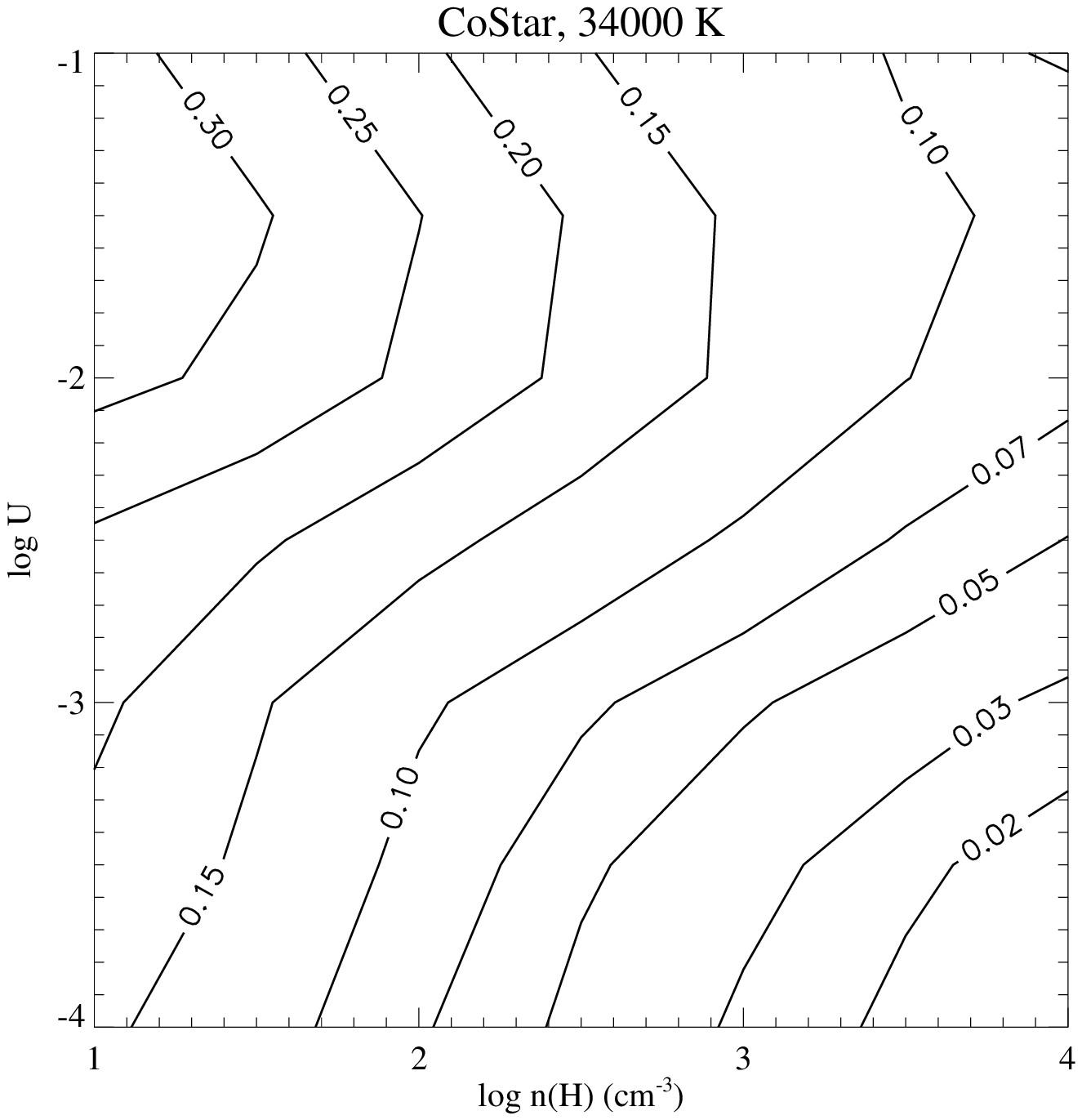}
\caption{Contribution to intensity of [C \footnotesize II\normalsize] 158 \micron\ from H \footnotesize II\normalsize\ region for WMBasic and CoStar atmospheres. The values of contour are the fraction of [C \footnotesize II\normalsize] 158 \micron\ intensity from H \footnotesize II\normalsize\ regions. For most part, the contribution from the H \footnotesize II\normalsize\ regions is between 3\% and 30\%. \label{fig5}}
\end{figure}

\begin{figure}
\includegraphics[scale=.45]{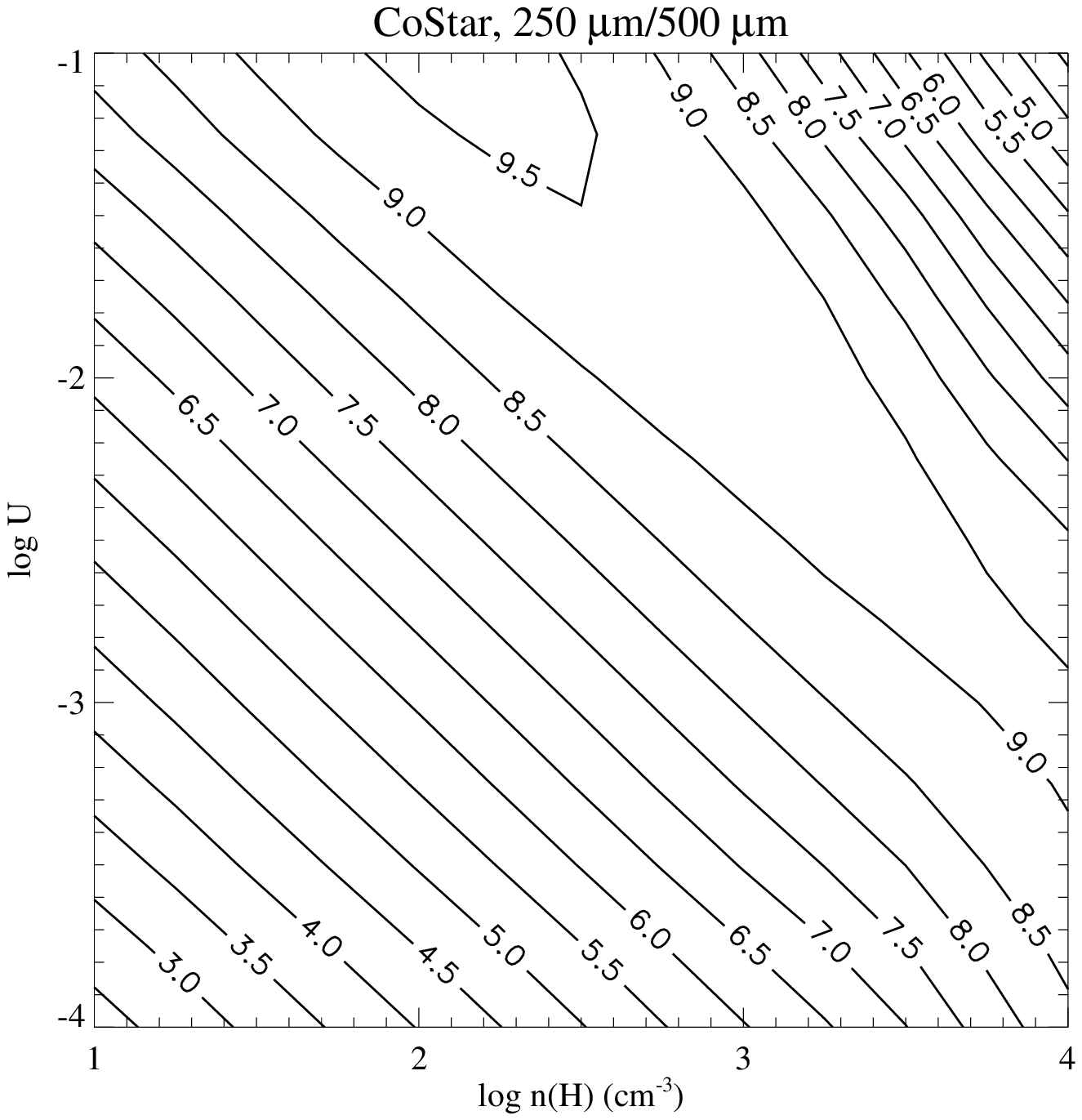}
\includegraphics[scale=.45]{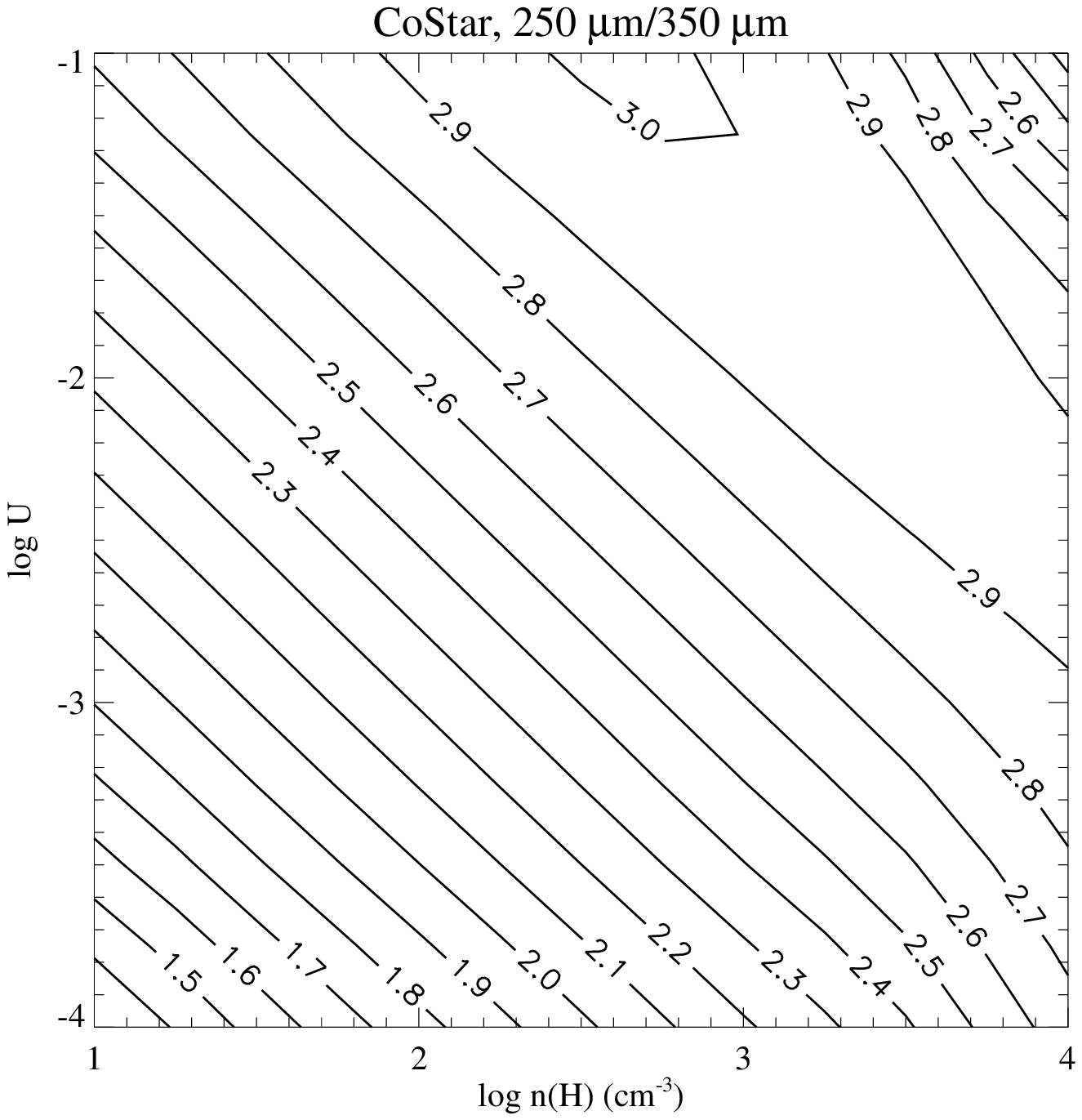}

\includegraphics[scale=.45]{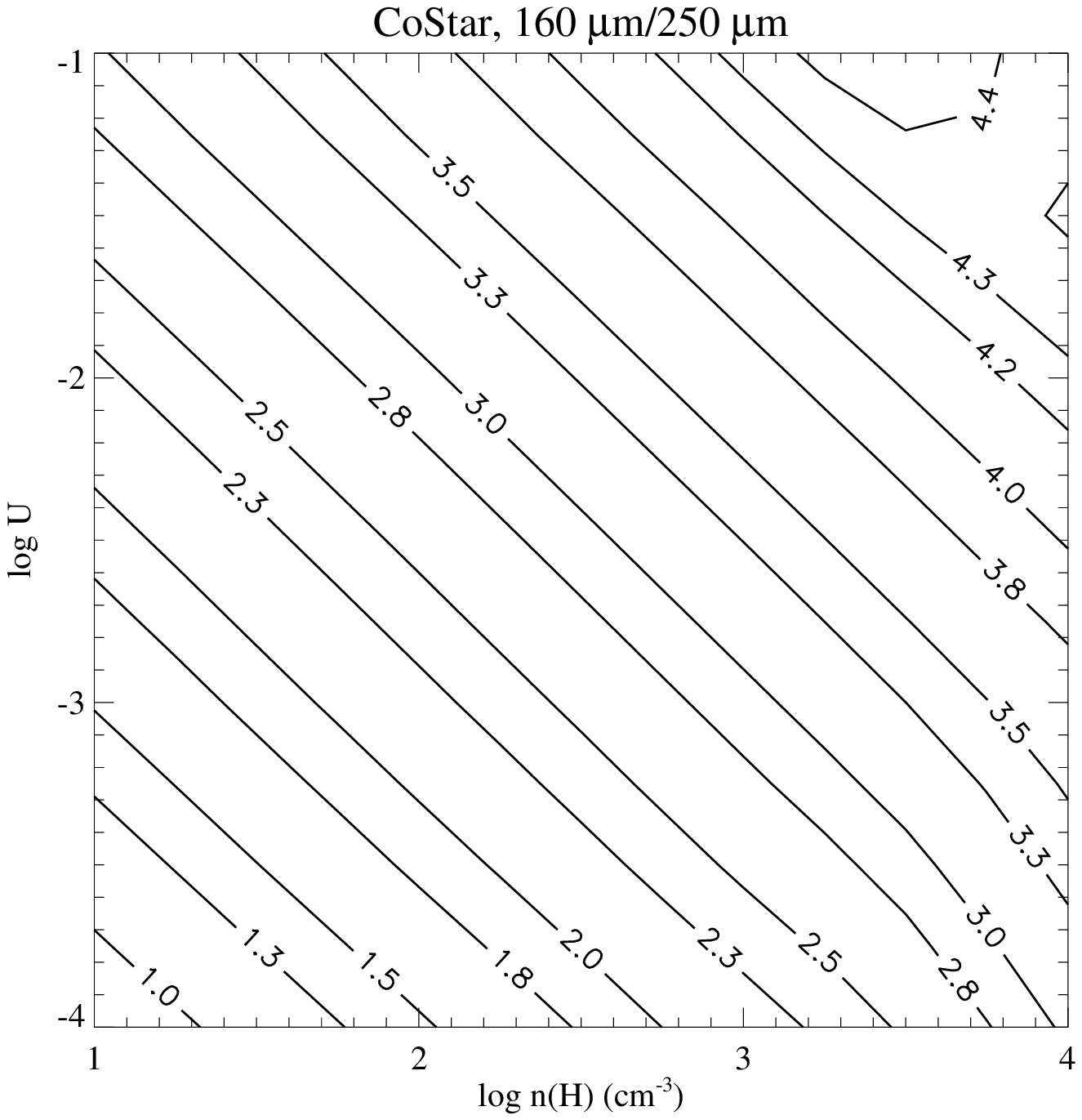}
\includegraphics[scale=.45]{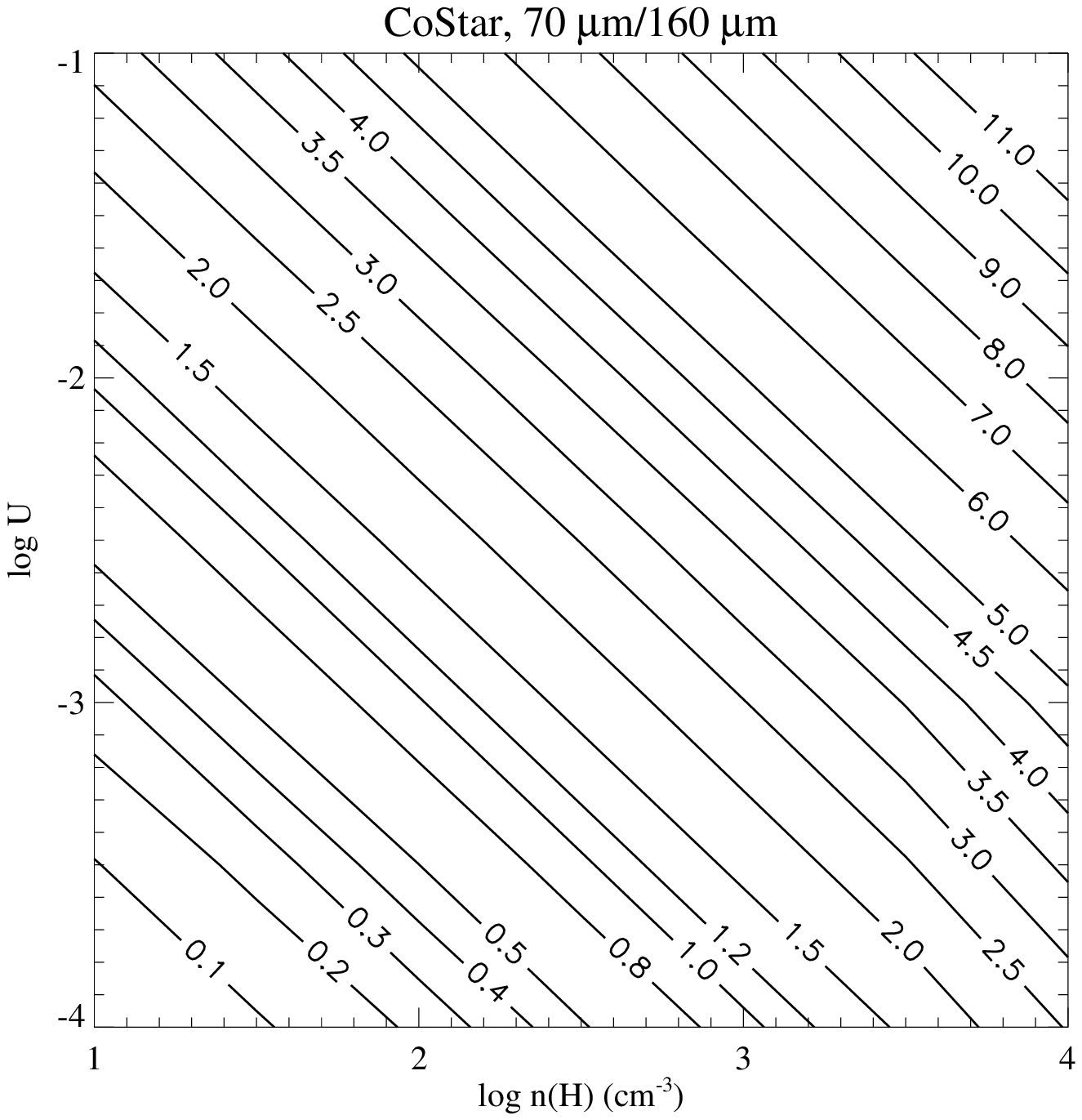}
\caption{Intensity ratios of FIR continuum for CoStar atmospheres at $T=38000$ K. \label{fig6}}
\end{figure}

\begin{figure}
\includegraphics[scale=.45]{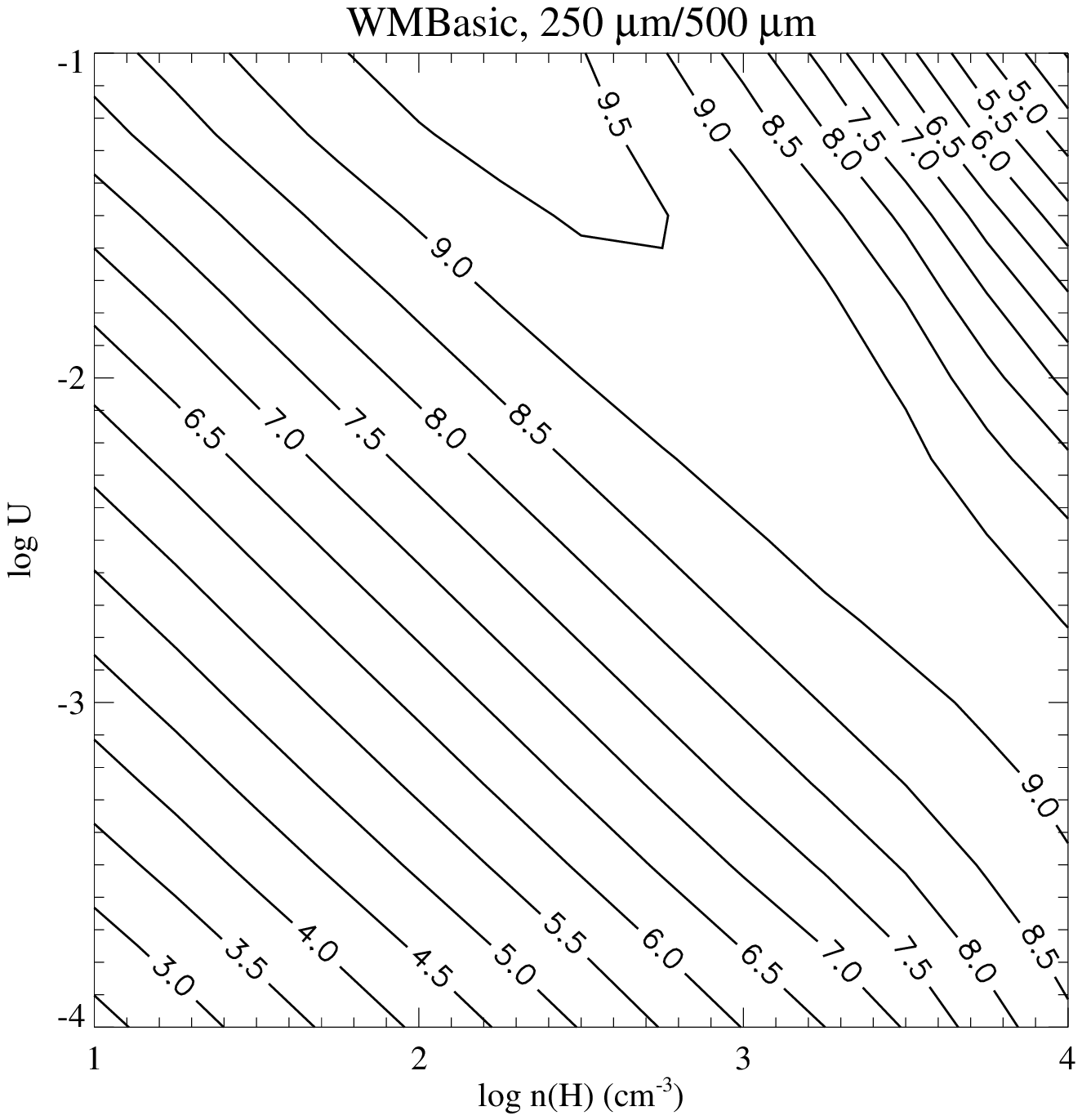}
\includegraphics[scale=.45]{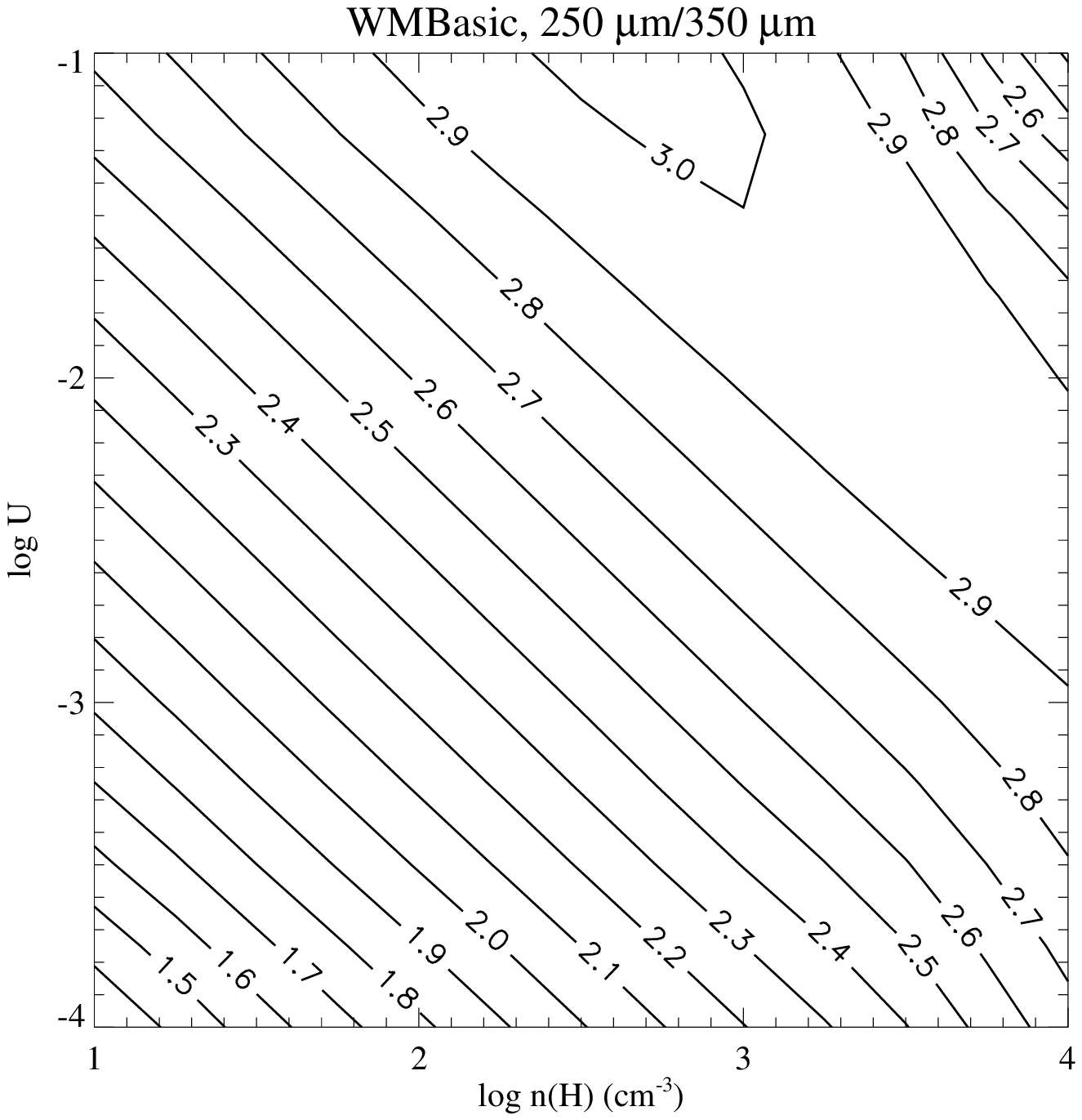}

\includegraphics[scale=.45]{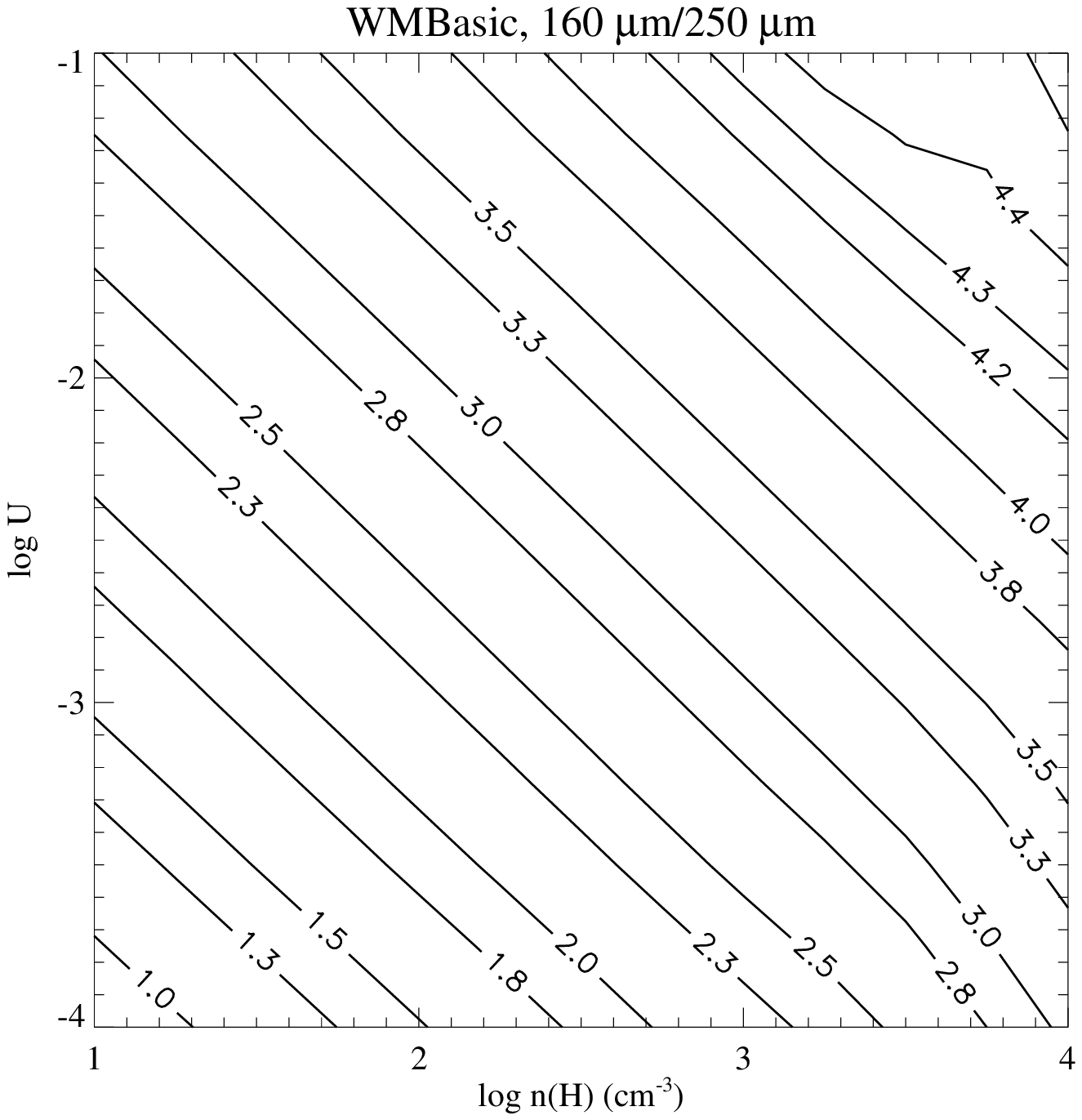}
\includegraphics[scale=.45]{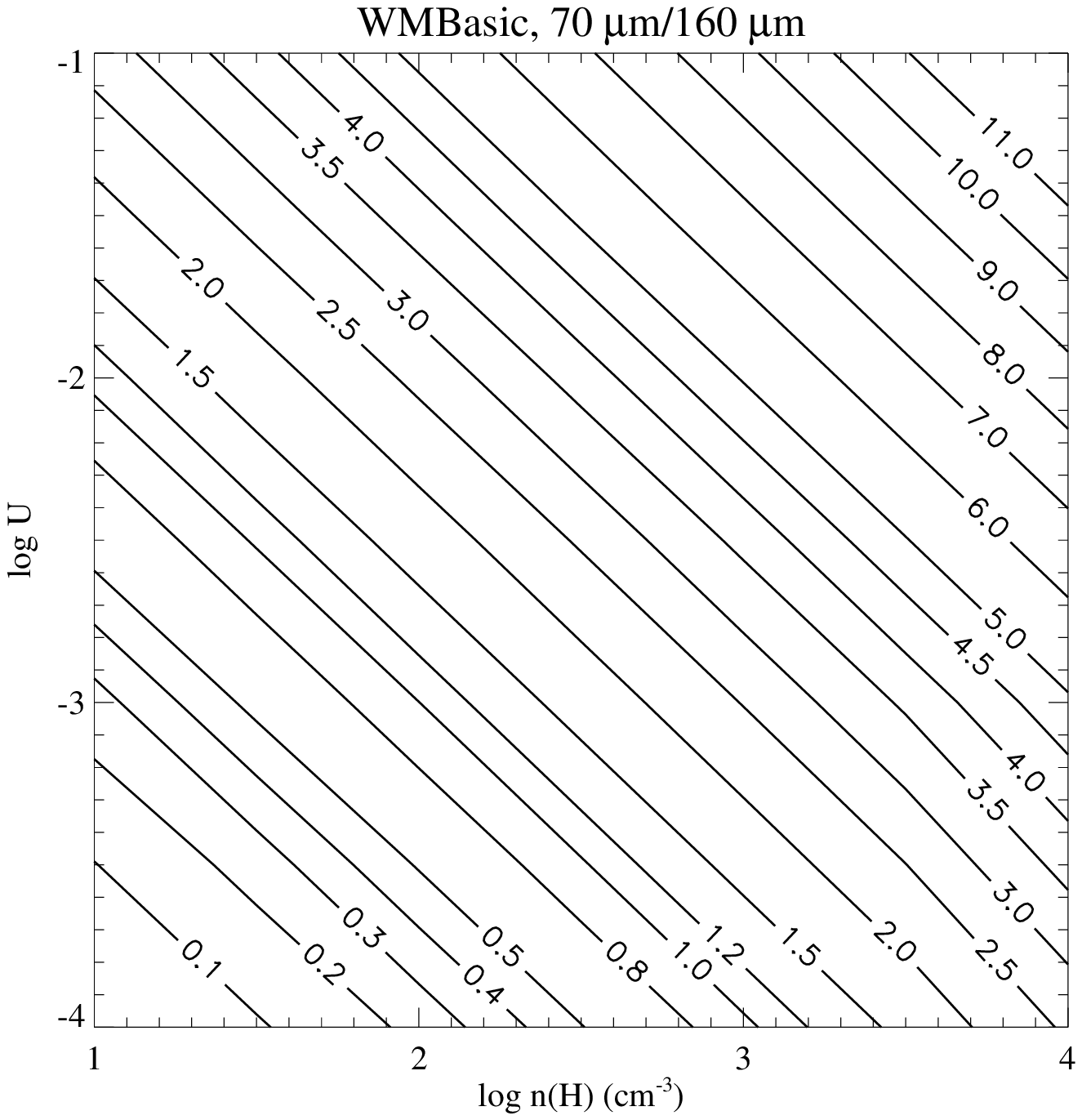}
\caption{Intensity ratios of FIR continuum for WMBasic atmospheres at $T=38000$ K. \label{fig7}}
\end{figure}

\begin{figure}
\includegraphics[scale=.42]{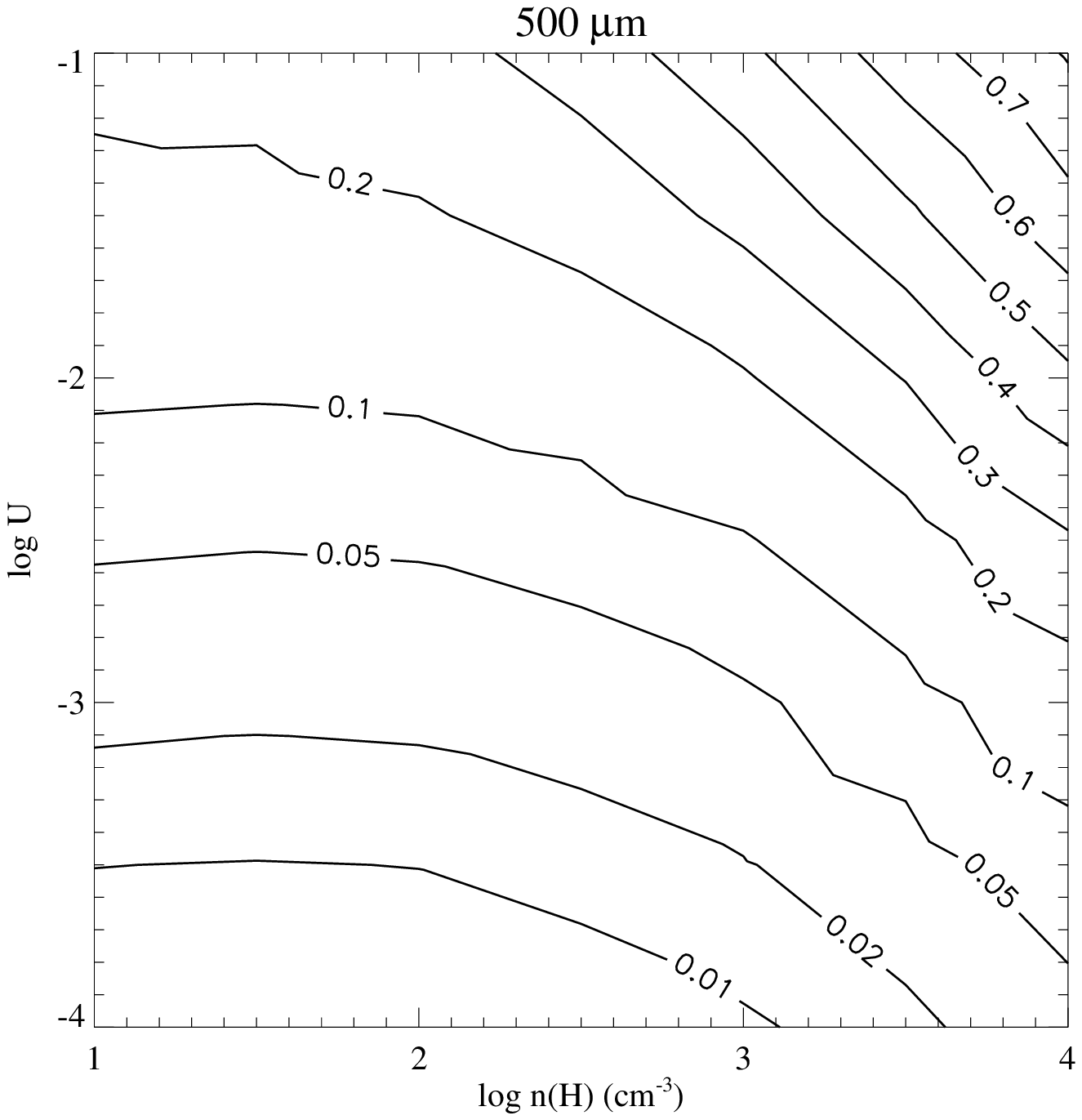}
\includegraphics[scale=.42]{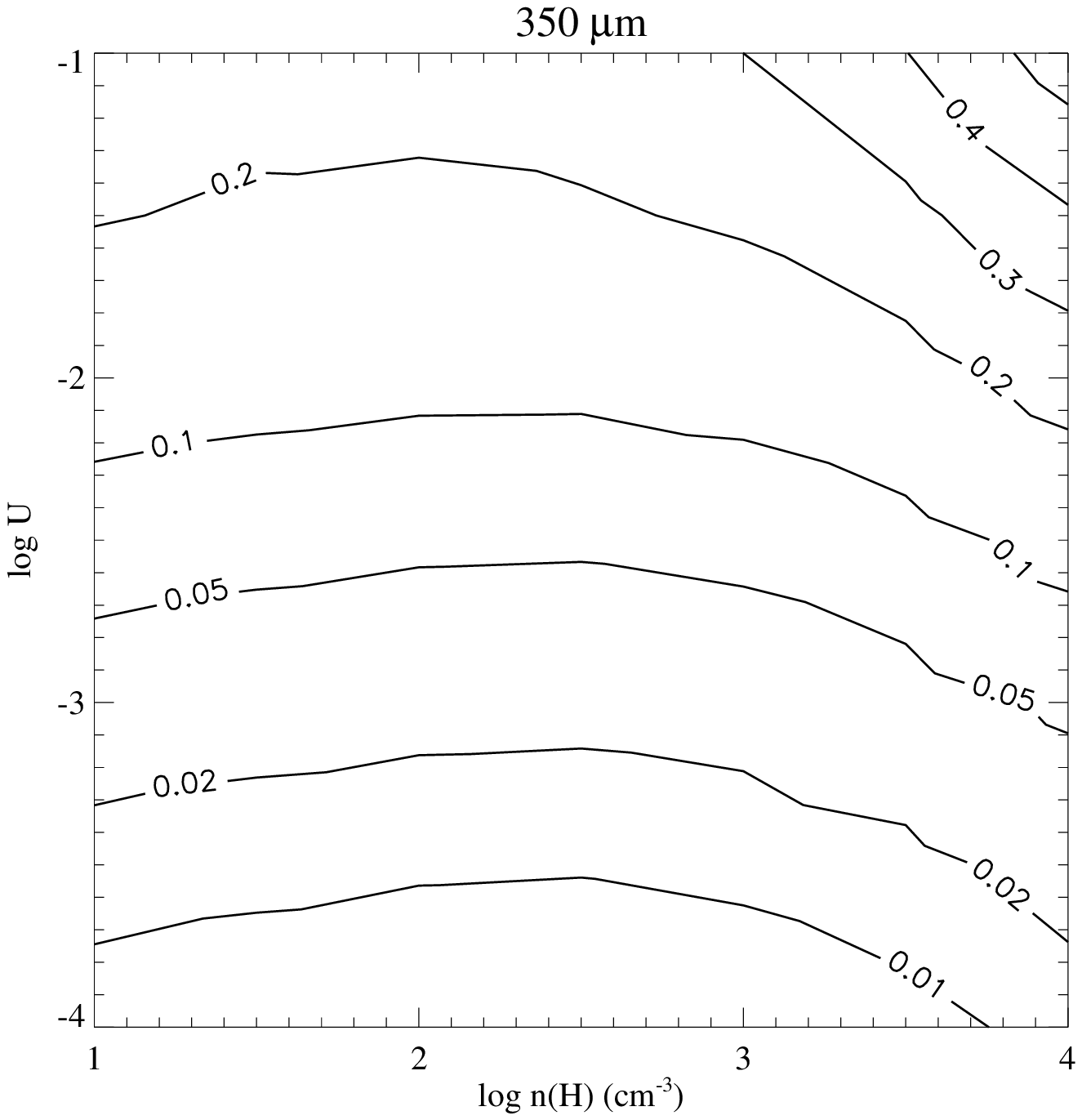}

\includegraphics[scale=.42]{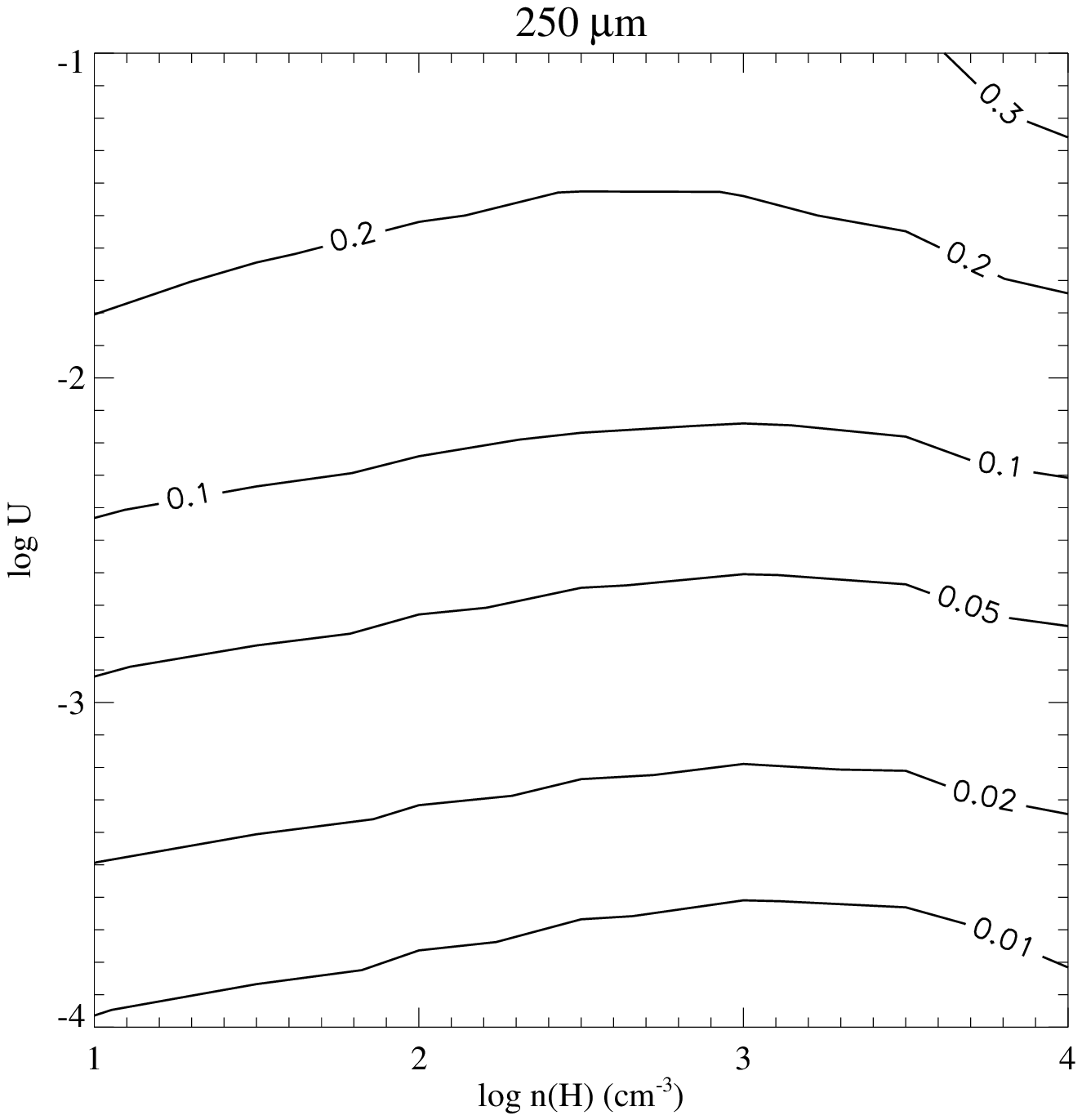}
\includegraphics[scale=.42]{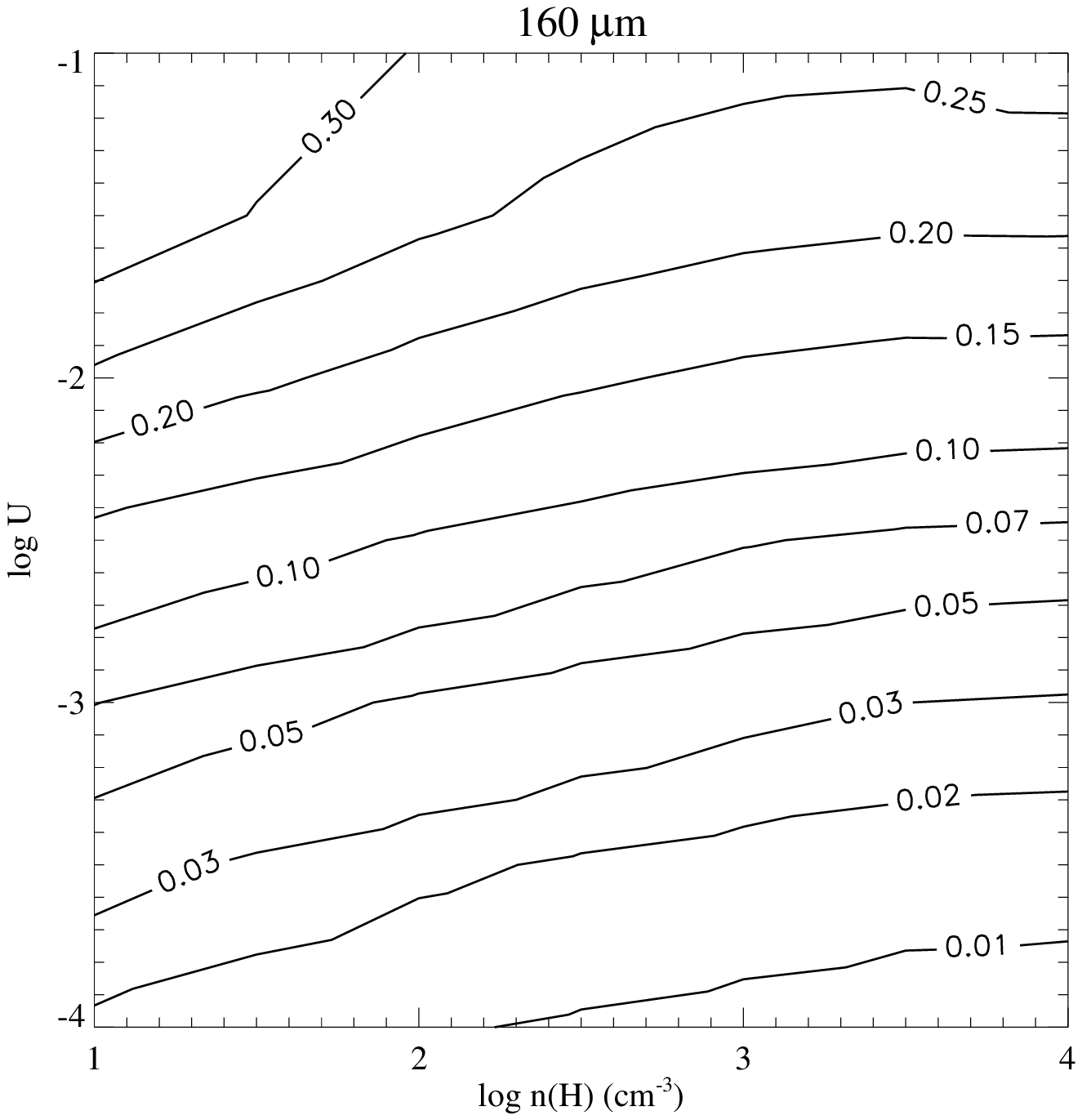}

\includegraphics[scale=.42]{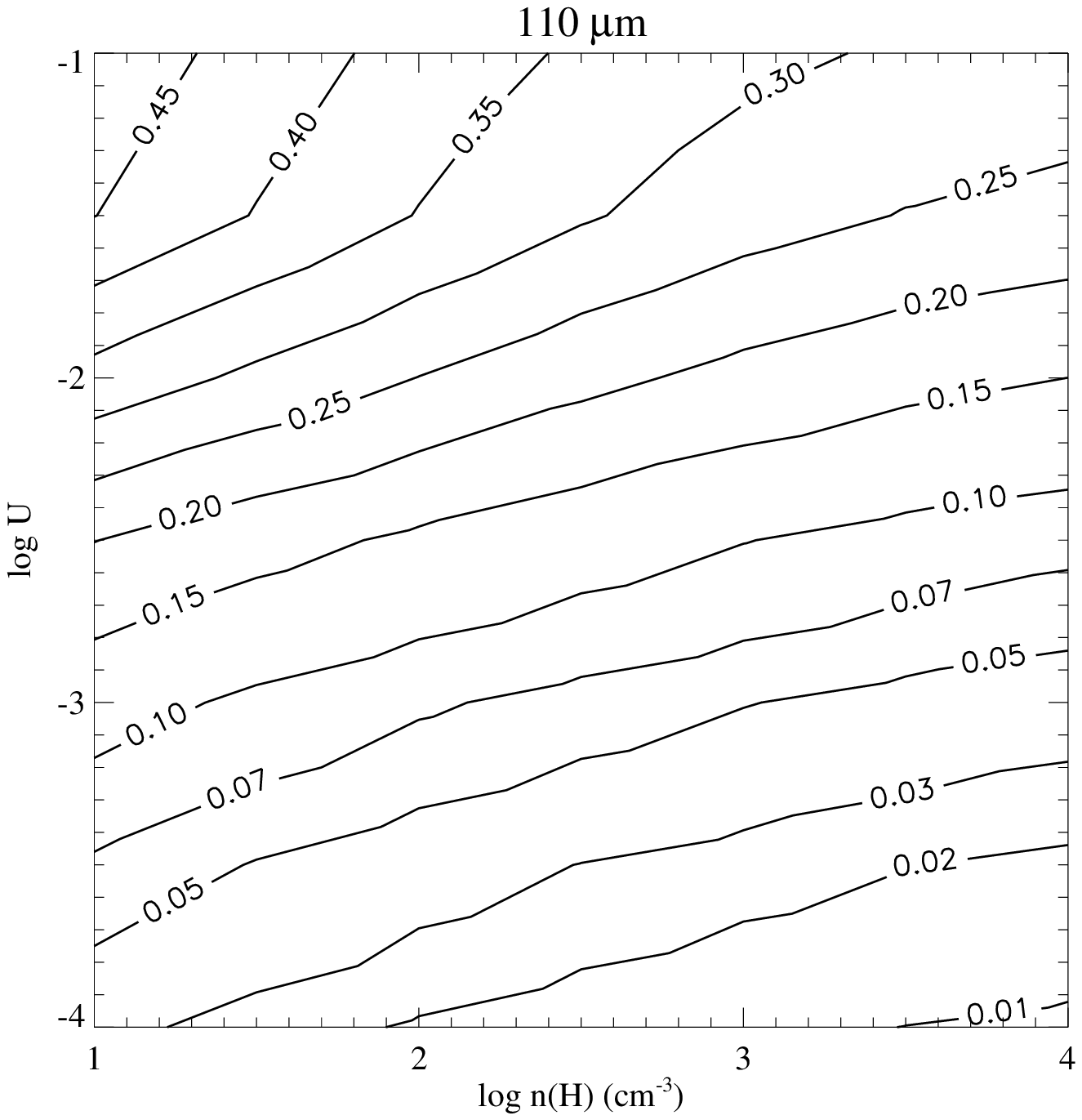}
\includegraphics[scale=.42]{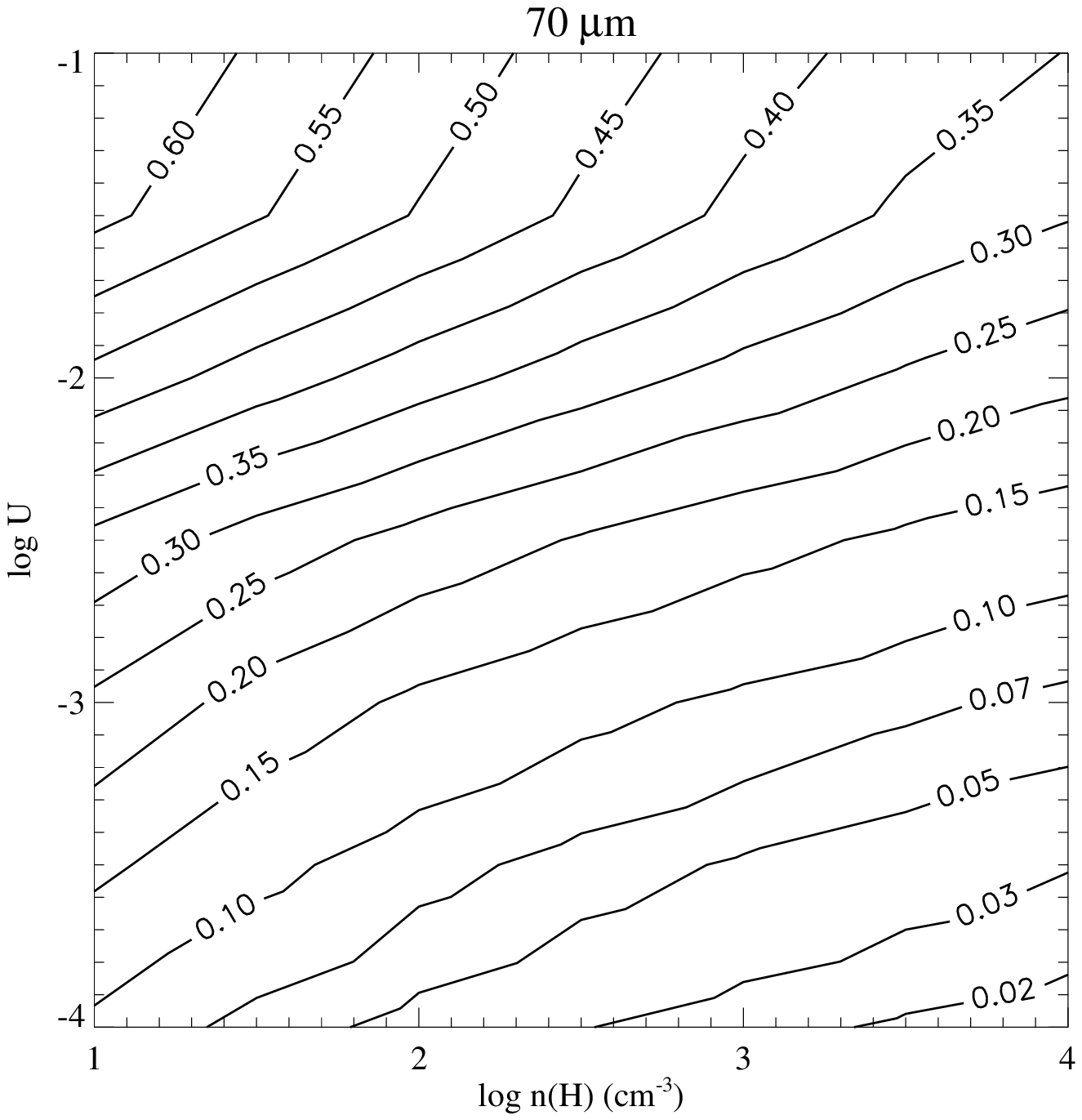}
\caption{Contribution to FIR continuum intensity of 70, 110, 160, 250, 350 and 500 \micron\ from H\footnotesize~II\normalsize\ regions for WMBasic atmospheres at $T=38000$ K. The value of contour level is the ratio of FIR continuum intensity from the H\footnotesize~II\normalsize\ region to its total intensity from the H\footnotesize~II\normalsize\ region and PDR. \label{fig8}}
\end{figure}

In this section, we compare our results with results of \citet{abel2005} and \citet{car94}, and discuss discrepancies. A significant gradient in [Ne \footnotesize III\normalsize ] 15.5 \micron/[Ne \footnotesize II\normalsize ] 12.8 \micron\ ratio is detected in NGC~253 \citep{dev2004}. The  [Ne \footnotesize III\normalsize ] 15.5 \micron/[Ne \footnotesize II\normalsize ] 12.8 \micron\ ratio is between 0.08 and 0.14 at the center region \citep{dev2004}, whereas \citet{ver2003} found it to be 0.07. The [S  \footnotesize IV\normalsize ] 10.5 \micron/[S  \footnotesize III\normalsize] 18.7 \micron\ ratio is about 0.03 \citep{ver2003}. To derive the value of $U$, we compare [Ne \footnotesize III\normalsize ] 15.5 \micron/[Ne \footnotesize II\normalsize ] 12.8 \micron\ ratio ($0.07\sim0.14$) and [S  \footnotesize IV\normalsize ] 10.5 \micron/[S  \footnotesize III\normalsize] 18.7 \micron\ ratio with Figure \ref{fig3} and \ref{fig4}. We find $\log U$ is $\sim$ -2 for CoStar atmospheres at $T=34000$ K. \citet{abel2005} found that the value of  [Ne \footnotesize III\normalsize ] 15.5 \micron/[Ne \footnotesize II\normalsize ] 12.8 \micron\ ratio and [S  \footnotesize IV\normalsize ] 10.5 \micron/[S  \footnotesize III\normalsize] 18.7 \micron\ ratio are increasing with effective temperature, and they are sensitive to $U$ and $T$. For WMBasic atmospheres, effective temperature 34000 K is not hot enough to produce [Ne\footnotesize~III\normalsize ]~15.5~\micron/[Ne \footnotesize II\normalsize ] 12.8~\micron\ ratio as large as 0.14 (Figure \ref{fig3}), and we find $\log U=-2.5$ for WMBasic atmospheres at 38000 K. The difference in effective temperature is caused by the discrepancy in SED shapes of stellar atmospheres (Figure \ref{fig1}). 

[S \footnotesize III\normalsize] 18.7 \micron/[S \footnotesize III\normalsize] 33.5 \micron\ ratio is $\sim0.5$ \citep{ver2003}. [O \footnotesize III\normalsize] 52 \micron/[O \footnotesize III\normalsize] 88 \micron\ ratio is $1\sim2$ \citep{car94}. Comparing those ratios with Figure 22 of \citet{abel2005}, we find that $n$(H) is between 100 cm$^{-3}$ and 200 cm$^{-3}$ for both stellar atmospheres. We adopt $n$(H)=$150$ cm$^{-3}$ as \citet{abel2005} did. Comparing the derived $U$ and density to other plots (Figure \ref{fig5} in this work and Figure 16, 17, 27, 29 and 33 of \citealt{abel2005}), we can deduce $G_0$, PDR density, line ratios and contributions for lines. 

We summarize all the predictions from our results, from \citet{abel2005}, and from \citet{car94} in Table \ref{tbl-1}. Our predictions of CoStar are consistent with \citet{abel2005}. Compared with CoStar atmospheres, our results for WMBasic atmospheres suggest $20\%$ less contribution from H \footnotesize II\normalsize\ region for [C \footnotesize II\normalsize] 158 \micron\ line intensity and 2.5 times greater for [Si \footnotesize II\normalsize] 35 \micron\ line intensity (Table \ref{tbl-1}). 

Both our calculations and \citet{abel2005} suggest a lower G$_0$ than what \citet{car94} deduced. Here we discuss this phenomenon qualitatively. To derive physical parameters including $G_0$, \citet{car94} performed two separate calculations, one for the H \footnotesize II\normalsize\ region and one for the PDR. They assumed $\sim30\%$ of the  [C \footnotesize II\normalsize] 158 \micron\ emission in NGC~253 originates in H \footnotesize II\normalsize\ regions, and assumed no [Si \footnotesize II\normalsize] 35 \micron\ emission in the PDR modeling. They followed the model of \citet{WTH90}, using the [C \footnotesize II\normalsize ] 158 \micron/[O \footnotesize I\normalsize ] 63 \micron\ intensity ratio and the line to continuum ratio, ($I$\footnotesize[Si II] 35 \micron\normalsize+$I$\footnotesize[O I] 63 \micron\normalsize+$I$\footnotesize[C II] 158 \micron\normalsize)/$I_{IR}$, to estimate $G_0$ and PDR density. Assuming that most [Si \footnotesize II\normalsize] 35 \micron\ emission comes from the H \footnotesize II\normalsize\ region, \citet{car94} adopted ($I$\footnotesize[O I] 63 \micron\normalsize+$I$\footnotesize[C II] 158 \micron\normalsize)/$I_{IR}$ in stead of ($I$\footnotesize[Si II] 35 \micron\normalsize+$I$\footnotesize[O I] 63 \micron\normalsize+$I$\footnotesize[C II] 158 \micron\normalsize)/$I_{IR}$ used in \citet{WTH90}. $G_0$ increases when [C \footnotesize II\normalsize] 158 \micron/[O \footnotesize I\normalsize] 63 \micron\ intensity ratio and line to continuum ratio decrease (Figure 1 in \citealt{WTH90}). For WMBasic atmospheres, our models predict only 20\% [C \footnotesize II\normalsize] 158 \micron\ arises from H \footnotesize II\normalsize\ regions, and as much as 50\% [Si \footnotesize II\normalsize] 35 \micron\ emission arises from PDR. For CoStar atmospheres, our models and \cite{abel2005} suggest as much as 80\% [Si \footnotesize II\normalsize] 35 \micron\ emission arises from PDR. Comparing with our predictions (Table \ref{tbl-1}), \citet{car94} overestimated the contribution from H \footnotesize II\normalsize\ regions to [C \footnotesize II\normalsize] 158 \micron\, and ignored the [Si \footnotesize II\normalsize] 35 \micron\ intensity from PDR, resulting in a lower [C \footnotesize II\normalsize] 158 \micron/[O \footnotesize I\normalsize] 63 \micron\ intensity ratio and a lower line to continuum ratio for PDR emission. Because of this underestimation of [C \footnotesize II\normalsize] 158 \micron/[O \footnotesize I\normalsize] 63 \micron\ ratio and line to continuum ratio, \citet{car94} derived a larger $G_0$ than predictions from our models and \citet{abel2005}. 

\subsection{Application to M82} 
M82 is a nearby ($D=3.25$ Mpc; \citealt{tam68}) starburst galaxy, well studied in FIR wavelength range (e.g., \citealt{lor96, col99}). In the core of M82, the active starburst region spans a diameter of 500 pc \citep{gri2001}.

H \footnotesize II\normalsize\ region diagnostics are observed at the center region of M82 \citep{bei2008}. [Ne \footnotesize III\normalsize] 15.5 \micron/[Ne \footnotesize II\normalsize] 12.8 \micron\ ratio is 0.15, and  [S \footnotesize IV\normalsize ] 10.5\micron/[S  \footnotesize III\normalsize] 18.7 \micron\ ratio is 0.036. Comparing observations with Figure \ref{fig3} and \ref{fig4}, we find that both ratios indicate $\log U$ to be $-2.5$ for WMBasic atmospheres at $T$ = 38000 K. Comparing Figure 22 of \citet{abel2005} with [O \footnotesize III\normalsize] 52 \micron/[O \footnotesize III\normalsize] 88 \micron\ ratio 1.24 \citep{col99} , we find density is to be 150 cm$^{-3}$. Using this $U$ and $n$(H), we derive other parameters with Figure \ref{fig5}, \ref{fig7} and \ref{fig8} in this work and Figure 16, 17, 27, 29 and 33 of \citet{abel2005}. Deduced parameters are summarized in Table \ref{tbl-2}.

In M82, dust continuum emission is strong in the superwind region and the very extended emission indicates dust distribution in the halo of this galaxy \citep{eng2006}. \citet{rou2010} found that FIR flux ratios would then be a natural consequence of the dilution of the radiation field with distance from the emitting stars.  The measured global flux densities are $457\pm2$ Jy at 250 \micron, $155\pm2$ Jy at 350 \micron, and $49.6\pm0.9$ at 500 \micron\ \citep{rou2010}. The center region of M82 has a complex structure, and the starburst emission from center region are 337, 111 and 35.4 Jy at 250, 350 and 500 \micron, respectively \citep{rou2010}. After subtracting the starburst emission from global fluxes, the values of 250~\micron/350~\micron\ ratio and 250~\micron/500~\micron\ ratio are 2.7 and 8.5. Contributions of FIR continuum intensity from H \footnotesize II\normalsize\ regions are similar for all three bands, at about 5\% (Figure \ref{fig8}). Compared with ratios of global emission and ratios of emission from wind and halo regions (Table \ref{tbl-2}), the predicted FIR continuum ratios (Figure \ref{fig7}) are more consistent with the wind and halo regions. The underestimation of 250~\micron/350~\micron\ ratio and 250~\micron/500~\micron\ ratio can be explained by the difference of dust size distribution between M82 and our assumption, since dust grains with different sizes emit FIR mission dominating FIR emission in different wavelength range. The predicted ratios of FIR continuum are more consistent with the FIR ratio for wind and halo regions suggests that size distribution of dust grains in these regions are closer to the model assumption than the center region. More detailed modeling exploration with Cloudy will be performed in the future to explain discrepancies between observations and models.

\section{SUMMARY}
The A05 model predicts diagnostic observables of an H \footnotesize II\normalsize\ region and an associated PDR based on two external parameters: the ionization parameter $U$ and initial total hydrogen density $n$(H) at the illuminated face, given the continuum shape and intensity of the ionization source, the chemical abundance of the gas, the condition of dust, and the geometry of the cloud. We explore the effect of different stellar atmospheres, discussing differences between calculation results for WMBasic and CoStar atmospheres. We presents the first set of plots of FIR ratios for Herschel bands,  and contributions to these specific FIR emission from H \footnotesize II\normalsize\ regions. 

[Ne \footnotesize III\normalsize ] 15.5 \micron/[Ne \footnotesize II\normalsize ] 12.8 \micron\ and [S  \footnotesize IV\normalsize ] 10.5 \micron/[S  \footnotesize III\normalsize] 18.7 \micron\ ratios are sensitive to stellar atmospheres. With the same $U$, WMBasic atmospheres need a higher stellar temperature than CoStar atmospheres to produce the same [Ne \footnotesize III\normalsize ] 15.5 \micron/[Ne \footnotesize II\normalsize ] 12.8 \micron\ and [S  \footnotesize IV\normalsize ] 10.5 \micron/[S  \footnotesize III\normalsize] 18.7 \micron\ ratios. 

We find that H \footnotesize II\normalsize\ regions can dominate the 500 \micron\ continuum emission when $\log U>-2$ and $\log n$(H)$>3$ cm$^{-3}$, and dominate the 70 \micron\ emission when $\log U>-2$ and $\log n$(H)$<2.5$ cm$^{-3}$. PDR is the main origin of 110, 160, 250 and 350 \micron\ continuum emission, although H \footnotesize II\normalsize\ regions still contribute more than 20\%  emission at $\log U>-1.5$.

We apply our results to two galaxies. For NGC~253, our results for CoStar atmospheres is consistent with \citet{abel2005}. Models for WMBasic atmospheres predict $\sim50\%$ of [Si \footnotesize II\normalsize] 35 \micron\ line intensity arises from the H \footnotesize II\normalsize\ region, while the H \footnotesize II\normalsize\ region only contributes $\sim20\%$ for CoStar atmospheres. For M82, we find G$_0\sim10^{2.8}$ and PDR density $\sim10^4$ cm$^{-3}$. The FIR continuum ratios predicted by model are more consistent with the wind and halo regions than the center regions of M82.

\acknowledgments
We thank Gary Ferland and Peter van Hoof, for their answers to our questions about Cloudy at on-line discussion forum\footnote{http://tech.groups.yahoo.com/group/cloudy\rule[-1.5pt]{0.2cm}{0.5pt}simulations/}. We thank the anonymous referee for helpful comments. This study is supported by funding KJCX2-YW-T20 of Chinese Academy of Science.

\clearpage

\clearpage

\clearpage

\clearpage

\clearpage

\clearpage

\clearpage

\clearpage

\begin{deluxetable}{lcccc}
\tabletypesize{\footnotesize}
\tablecaption{Application to NGC~253\label{tbl-1}}
\tablewidth{0pt}
\tablehead{
\colhead{Parameters} &\colhead{CoStar} &\colhead{WMBasic} &\colhead{Abel et al. (2005)} &  \colhead{Carral et al. (1994)}
}
\startdata
$\log U $&-2& -2.5 & -2 & -\\
$n$(H) (cm$^{-3}$)&150&150&150&$430^{+290}_{-225}$\\
$G_0$&5E3&1E2.8&5E3&2E4\tablenotemark{a}\\
PDR density (cm$^{-3}$)&2E3$\sim$2E4&2E3$\sim$4E4&2E3$\sim$2E4&1E4\tablenotemark{a}\\
$I$\tiny[O I] 63 \micron\footnotesize/$I$\tiny[C II] 158 \micron\footnotesize&1&1&$1$&$0.8\sim1.1$\\
Contribution to [C \tiny II\footnotesize]\tablenotemark{b}&25\%&20\%&30\%&30\%\\
Contribution to [Si \tiny II\footnotesize]\tablenotemark{b}&20\%&50\%&20\%&-\\
\enddata
\tablenotetext{a}{Accurate to about a factor of 2.}
\tablenotetext{b}{The percentage of this line intensity from H \tiny II\footnotesize\ regions.}
\end{deluxetable}

\begin{deluxetable}{lccl}
\tabletypesize{\footnotesize}
\tablecaption{Application to M82\label{tbl-2}}
\tablewidth{0pt}
\tablehead{
\colhead{Parameters} & \colhead{Observations} & \colhead{Model Predictions} &\colhead{References} 
}
\startdata
$\log U $&$-3.5$; $-2.5$&$-2.5$&1, 2\\
$n$(H) (cm$^{-3}$)&$250$; $100$&$150$&1, 2\\
$G_0$&$10^{2.8}$&$10^{2.8}$&1\\
PDR density (cm$^{-3}$)&$10^{3.3}$; $10^4\sim10^5$&$10^{3.3}\sim10^{4.6}$&1, 2\\
$I$\tiny[O I] 63 \micron\footnotesize/$I$\tiny[C II] 158 \micron\footnotesize&$1.38\pm0.03$&1&1\\
Contribution to intensity of [C \tiny II\footnotesize] 158 \micron \tablenotemark{a}&$25\%$&$20\%$&1\\
$I_{250~\micron}/I_{500~\micron}$\tablenotemark{b}&9.2\tablenotemark{c}; 8.5\tablenotemark{d}&$7.5$&3\\
$I_{250~\micron}/I_{350~\micron}$\tablenotemark{b}&2.9\tablenotemark{c}; 2.7\tablenotemark{d}&$2.5$&3\\
\enddata
\tablenotetext{a}{The percentage of this line intensity from H \tiny II\footnotesize\ regions.}
\tablenotetext{b}{Ratio of FIR continuum intensity at different wavelength.}
\tablenotetext{c}{Ratio of global flux.}
\tablenotetext{d}{Ratio of wind and halo flux.}
\tablerefs{(1) \citet{col99}; (2) \citet{spi92}; (3) \citet{rou2010}.}
\end{deluxetable}

\end{document}